\documentclass[11pt, A4]{article}
\usepackage{slashbox}
 
\usepackage{bbm}
\usepackage{bbold}

\usepackage{amsfonts}

 \usepackage{amsmath,amssymb,cite,graphicx,multirow}

\usepackage{cancel}
\usepackage{cite}

\usepackage{graphicx}
\usepackage{epstopdf}
\usepackage{amsmath}
\usepackage{amssymb}
\usepackage{amsthm}
\usepackage[usenames]{color}
\usepackage{array}
\usepackage{float} 

\usepackage[
      colorlinks=true,
      linkcolor=blue,
      urlcolor=blue,
      filecolor=blue,
      citecolor=red,
      pdfstartview=FitV,
      pdftitle={},
      pdfauthor={},
      pdfsubject={},
      pdfkeywords={},
      pdfpagemode=None,
      bookmarksopen=true
]{hyperref}

\usepackage{epsfig}
\usepackage{mathtools}

\usepackage{comment}

\def\ZZZ{\mathbbm{Z}}
\def\zzz{\mathbbm{Z}}

\def\half{\frac{1}{2}}

\def\be{\begin{equation}}
\def\ee{\end{equation}}
\def\bea{\begin{eqnarray}}
\def\eea{\end{eqnarray}}

\newcommand{\rom}[1]{\mathrm{#1}}

\def\nn{\nonumber}

\begin{document}

  \begin{centering}

{\flushright {Preprint 2022}\\[15mm]}

{\LARGE \textbf{ Positivity of discrete information \\   \vspace{0.3cm} for  CHL black holes} } \\

 \vspace{0.8cm}
{
Suresh Govindarajan$^a$, Sutapa Samanta$^b$,
P Shanmugapriya$^c$,
and Amitabh Virmani$^c$}
\vspace{0.8cm}

\begin{minipage}{.9\textwidth}  \small
\begin{center}
$^a$Department of Physics,  Indian Institute of Technology Madras, \\
Chennai,  India 600036 \\
  \vspace{0.5cm}
$^b$Department of Physics and Astronomy, Western  Washington University, \\
516 High Street, Bellingham, WA 98225\\
  \vspace{0.5cm}
$^c$Chennai Mathematical Institute, H1 SIPCOT IT Park, \\ 
Kelambakkam, Tamil Nadu, India 603103 \\
  \vspace{0.5cm}
{\tt suresh@physics.iitm.ac.in, samants2@wwu.edu, \{shanmugapriya, avirmani\}@cmi.ac.in}\\ 
  \vspace{0.5cm}
\end{center}
\end{minipage}

\end{centering}

\begin{abstract}

Black holes carry more information about the microstates than just the total degeneracy.  As a concrete example, the $\ZZZ_N$-twined helicity trace indices for $\frac{1}{4}$-BPS black holes of 
the CHL models allow  extracting information about the distribution of the $\ZZZ_N$ charges among the black hole microstates.  The number of black hole microstates carrying   a definite eigenvalue under the generator of the $\ZZZ_N$ twining group must be positive.  This leads to a specific prediction for the signs of certain linear combinations of Fourier coefficients of 
Siegel modular forms. We explicitly test these predictions for low charges.
 In the D1-D5-P duality frame, we compute the appropriate hair removed partition functions and show the positivity of the appropriate Fourier coefficients for low charges. We present various consistency checks on our computations. 
 \end{abstract}

\newpage

\tableofcontents

\setcounter{equation}{0}
\section{Introduction} 

We now have a fairly detailed understanding of the degeneracy  of states that contribute to the entropy of $\frac{1}{4}$-BPS black holes in $\mathcal{N}=4$ supersymmetric string theories \cite{Sen:2007qy, Mandal:2010cj}. Specifically, the CHL models \cite{Chaudhuri:1995fk, Chaudhuri:1995bf, Schwarz:1995bj, Chaudhuri:1995dj}  in four dimensions provide a rich playground for studying the physics of  BPS black holes \cite{hep-th/0503217, hep-th/0505094, hep-th/0510147, hep-th/0602254, hep-th/0605210, hep-th/0609109, Dabholkar:2007vk,  Cheng:2007ch, Dabholkar:2008zy, Banerjee:2009uk, Jatkar:2009yd, 0907.1410}.
Many detailed agreements between the microscopic and macroscopic sides have been established.  In almost all these calculations, one computes an index rather than absolute degeneracy on the microscopic side. The index is defined so that it only receives contributions from the BPS states preserving the right amount of supersymmetry.  The index so-defined  is also protected; it does not change as we vary the moduli of the theory.

In four dimensions, the ideal indices that capture the protected information are the helicity trace indices~\cite{Sen:2007qy, Mandal:2010cj}.  The helicity trace index relevant for $\frac{1}{4}$-BPS black holes in $\mathcal{N}=4$ supersymmetric string theories is defined as
\be \label{b6}
B_{6} = {1\over 6!}\,
\mathrm{Tr} \left[ (-1)^{2h} (2h)^{6}\right],
\ee
where the trace is taken over all states
carrying a given set of charges, and where $h$ is the third component of the angular momentum of
a state in the rest frame. For spherically symmetric four-dimensional supersymmetric black holes $(-1)^{2h}=1$
 \cite{0903.1477}. As a result, the helicity trace index $B_{6}$ \eqref{b6} is directly related to the absolute degeneracy.

 If the theory admits an additional  discrete symmetry, and if we further  restrict ourselves to dyonic states with charges that are invariant under the action of the discrete symmetry, we can define twined helicity trace indices \cite{0911.1563, 1002.3857, 1006.3472}. These indices  capture more refined information about the black hole microstates.

\subsection*{Discrete information}

The twined helicity trace index 
is defined as
\be
B^{g_N}_{6}= {1\over 6!}\,
\mathrm{Tr}\left[ g_N\, (-1)^{2h} (2h)^{6}\right],
\ee
where $g_N$ is the generator of the discrete symmetry of order $N$. We will be mostly concerned with 
$B^{g_N}_{6}$ for a class of type II CHL models. 
We choose $g_N$ to be the generator of a geometric $\ZZZ_N$ acting on K3 or on one of its CHL orbifolds K3$/\ZZZ_M$; and require $g_N$ to commute with all 16 unbroken supersymmetries of the $\ZZZ_M$ twisted CHL compactification. 
Such indices are often called twisted-twining indices.

 It is now well understood  \cite{0911.1563, 1002.3857, 1006.3472, 1312.0622} that such a twisted-twining index is given by the Fourier coefficients of the twisted-twining black hole partition function---the inverse of a Siegel modular form of a subgroup of Sp$(2, \ZZZ)$. Various properties of such twisted-twining partition functions from both the microscopic and macroscopic sides have been studied \cite{Cheng:2010pq, Chowdhury:2014lza, Govindarajan:2019ezd, Govindarajan:2020owu}.

In this paper, our main interest is in the index 
\be 
S_a = {1\over 6!} \mathrm{Tr}_a \left[ (-1)^{2h}(2h)^6\right],
\ee 
for states carrying a \emph{definite} $g_N$ eigenvalue $e^{2\pi i a/N}$ with $0 \le a \le N-1$. 
This index is closely related to $B^{g_N}_{6}$. To obtain $S_a$ we first repeat the
analysis 
of 
$B^{g_N}_{6}$
with $g_N$ replaced by $(g_N)^b$ for any integer $b$. 
The role of $N$ is now played by the order of $(g_N)^b$. This allows
us to compute $S_a$ via  the discrete Fourier transform,
\be \label{linear_comb}
S_a =  {1\over 6!} {1\over N}\, \sum_{b=0}^{N-1} e^{-2\pi i ab/N} \, 
\mathrm{Tr}\left[ (g_N)^b(-1)^{2h} (2h)^6 \right].
\ee

We mentioned around \eqref{b6} that the index $B_6$ is directly related to the absolute degeneracy. In a duality frame, where hair modes are only the fermion zero modes, the exact relation is \cite{1008.4209},
\be
- B_6 = d_\rom{hor}. \label{equiv}
\ee All the elements that go into the argument that leads to \eqref{equiv} remain valid even with the $g$ insertion \cite{0911.1563}: it is for the same  spherically symmetric attractor geometry that the indices $B_6^g$ are being computed. Thus, $B_6^g$ is simply negative of the \emph{absolute degeneracy weighted with $g$.} Since a degeneracy must be a \emph{positive integer}, this leads to a specific prediction for the signs of linear combinations \eqref{linear_comb} of the Fourier coefficients of the Siegel modular forms that capture the indices $B_6^{g_N^b}$. In this paper, we will be computing  $B^{g^b_N}_{6}$ for $ 0\le b \le N-1$  for the various $\ZZZ_M$ CHL models with additional discrete $\ZZZ_N$ twining symmetry and study the positivity properties of $S_a$ for $ 0\le a \le N-1$.

For example, for $N=2$  we have the index $S_0$ for states  with   eigenvalue +1 under $g_2$ to be
\be
S_0 = \frac{1}{2} \left( B_{6} + B^{g_2}_{6} \right),
\ee
and the index $S_1$ for states  with  eigenvalue $-1$  under $g_2$ to be
\be
S_1 = \frac{1}{2} \left( B_{6} - B^{g_2}_{6} \right).
\ee
This implies, $B_6 = S_0 + S_1$ and $ B^{g_2}_{6} = S_0 - S_1$. Upto an overall sign,  $B_6$ simply counts the total number of states disregarding any information the state may carry about the $\ZZZ_2$ charge. On the other hand, $ B^{g_2}_{6}$ counts the number of states weighted with the $\ZZZ_2$ charge. If the number of states with $+1$   eigenvalue and $-1$   eigenvalue are roughly the same, then  we expect $S_0 \approx S_1$ and the twined indices $ B^{g_2}_{6} $ to be smaller (positive or negative) numbers compared to $B_6$. Indeed,  we will see this in later sections; and is also expected from the AdS$_2$/CFT$_1$ interpretation of the twined indices \cite{1002.3857}.\footnote{In the AdS$_2$/CFT$_1$ interpretation, the  computation of  the twisted-twining  indices can be expressed as a path integral with a suitable AdS$_2$ asymptotics. Twining requires a $g_N$-twisted boundary condition on the fields in carrying out such a path integral. The saddle that contributes to the $\ZZZ_N$ twined partition function is a  $\ZZZ_N$ orbifolds of $AdS_2 \times S^2$.}

Similarly, for $N=3$  we have the number of states $S_0$ (upto an overall sign) with  $+1$ eigenvalue under $g_3$ to be
\be
S_0 = \frac{1}{3} \left( B_{6} + B^{g_3}_{6}+ B^{g^2_3}_{6} \right),
\ee
and the number of states $S_1$ and $S_2$ with  eigenvalues $\omega = e^{2\pi i /3}$  and $e^{4\pi i /3}$ respectively under $g_3$   to be
\bea
&& S_1 = \frac{1}{3} \left( B_{6} + \omega^{-1} B^{g_3}_{6}+\omega^{-2} B^{g^2_3}_{6} \right),\\
&& S_2 = \frac{1}{3} \left( B_{6} + \omega^{-2} B^{g_3}_{6}+\omega^{-1} B^{g^2_3}_{6} \right).
\eea

\subsection*{Hair removal} 

If two black holes have identical near-horizon geometries, they must have identical microscopic indices. There is, however, a well-studied apparent ``counterexample'' to this: the BMPV black hole in flat space~\cite{Breckenridge:1996is} versus the BMPV black hole in Taub-NUT space~\cite{Gauntlett:2002nw, hep-th/0503217}.   These black holes have identical near-horizon geometries but different  microscopic indices. It is now well understood that  the key to the resolution of this puzzle is the black hole hair modes \cite{Banerjee:2009uk, Jatkar:2009yd, Chakrabarti:2020ugm, Chattopadhyaya:2020yzl}: smooth, normalisable,  bosonic and fermionic degrees of freedom living outside the horizon.  For the case of $\text{K3}$ compactification of type IIB theory, Sen et al  constructed hair modes as non-linear solutions to the supergravity equations\cite{Jatkar:2009yd}. They showed that once the contributions of the hair modes are removed, the 4d and 5d partition functions match. This analysis was recently extended to CHL models \cite{Chakrabarti:2020ugm, Chattopadhyaya:2020yzl}. Given these results, it is fairly clear that the twisted-twining hair removed partition functions also match in 4d and 5d.

In this paper, we also study the construction of the 4d hair removed partition functions and positivity properties of the corresponding Fourier coefficients. More precisely, we will construct the analogs of $S_a$ from the hair removed partition functions in the D1-D5 duality frame.

\subsection*{Organisation}

The rest of the paper is organised as follows.  
In section \ref{sec:twining}, we consider twining black hole hair removal for four-dimensional BMPV black hole. For different $\mathbbm{Z}_{N}$ twinings, we construct the 4d hair partition functions.  
In section \ref{sec:positivity_twining}, we study the positivity of the coefficients $S_a$ defined in equation \eqref{linear_comb} for $\mathbbm{Z}_{N}$ twinings $N=2, 3, 4$.  These three cases are representative of more general cases and capture the essential ideas. We compute the appropriate Fourier coefficients from the full 4d partition functions and also from the hair removed 4d partition functions. In the D1-D5 frame the hair removed partition functions capture the horizon states. In a duality frame with no hair (other than the fermion zero modes), the full 4d partition functions capture the horizon states.

In section \ref{sec:twisted_twining}, we consider twisted-twining black hole hair removal. For different $\mathbbm{Z}_{M}~\times~\mathbbm{Z}_{N}$ models, we construct the 4d hair partition functions.  
In section \ref{sec:positivity_twisted_twining}, we study the positivity of the coefficients $S_0$ and $S_1$
for the $\ZZZ_2~\times~\ZZZ_2$ CHL model. In turns out that multiple representations for the twisted-twining partition function for this model have been proposed in the literature, but the equivalence of these representations has not been formally established.  We compute the coefficients $S_0$ and $S_1$
using different representations and obtain identical results. 
In section \ref{sec:other_models}, we study the positivity of coefficients $S_a$ for some other models. 
We close with a brief discussion in section \ref{sec:disc}.

\subsection*{Other studies}
 
 Positivity of the Fourier coefficients without twining for CHL models was first explored in \cite{1008.4209}.  A proof of  this positivity property for a special class of four and five dimensional black holes in the unorbifolded model was presented in \cite{Bringmann:2012zr}. Positivity of the  Fourier coefficients of the hair removed partition functions without twining for CHL models was explored in~\cite{Chattopadhyaya:2017ews, Chattopadhyaya:2018xvg, Chattopadhyaya:2020yzl}.

\setcounter{equation}{0}

\section{Twining black hole hair removal}
\label{sec:twining}
We begin by considering type IIB string theory compactified on K3$~\times~\mathrm{S}^1~\times~\widetilde{\mathrm{S}}^1$. In a subspace of the moduli space of this compactification, we identify $g_N$ to be the generator of a specific geometric $\mathbbm{Z}_{N}$ symmetry of K3 that preserves all the covariantly constant spinors of K3 and leaves invariant some 2-cycles of K3.  In such a compactification, we consider the D1-D5-P-KK system preserving 4 of the 16 supersymmetries as follows \cite{Sen:2007qy}:  a single D5 brane wrapped on K3 $\times~ \mathrm{S}^1$, $Q_1$ D1-branes wrapped on S$^1$, a single KK monopole with negative charge associated with the circle $\widetilde{\mathrm{S}}^1$, left moving momentum $-n$ along S$^1$, and right moving momentum $J$ along  $\widetilde{\mathrm{S}}^1$. Since the D5 brane wraps K3,  it also carries a negative D1 charge \cite{Bershadsky:1995qy}. The net D1 charge is therefore, $Q_1 - 1$. 
For such a set-up,  T-duality invariant charge bilinears are
\begin{align} \label{bilinears}
Q^2 &= 2n, & P^2 &= 2 (Q_1 - 1), & Q \cdot P = J.
\end{align}
We will write most of our formulae below we in terms of the T-duality invariants $Q^2$, $P^2$, and $Q \cdot P$. However, with regard to the hair removal discussions, it is best to keep the above brane configuration in mind. 

For this set-up, in the region of the moduli space where the type IIB string coupling is small, the result for the twined index $B^{g_N}_{6}$ is~\cite{0911.1563}
\be\label{B6gNg}
-B^{g_N}_{6}  = (-1)^{Q\cdot P+1}\,
g\left({1\over 2} Q^2 , {1\over 2}\, P^2,
Q\cdot P\right)\, ,
\ee
where $g(l, k,j)$ are
the coefficients of Fourier expansion of the function
$1/ \widetilde \Phi(\widetilde \rho,\widetilde \sigma, \widetilde v)$:
\be 
{1
\over \widetilde\Phi(\widetilde \rho,\widetilde \sigma, \widetilde v)}
=\sum_{l,k,j} g(l,k,j) \, e^{2\pi i (l\, \widetilde \rho + k\,
\widetilde\sigma
+ j\, \widetilde v)}\, . \label{fourier_expansion}
\ee
The function\footnote{The  widetilde notation on modular forms and on the coordinates of the Siegel upper half space is somewhat standard in the CHL literature \cite{Sen:2007qy}.}
  $\widetilde \Phi(\widetilde \rho,\widetilde \sigma, \widetilde v)$ is:
\bea
\widetilde{\Phi}(\widetilde \rho, \widetilde \sigma,\widetilde v) &=&  e^{2\pi i 
\left(\widetilde \rho+\widetilde \sigma+ \widetilde v\right) } \nonumber \\
&& \prod_{b=0}^1\,
\prod_{r=0}^{N-1}\,  \prod_{(k,l)\in \mathbbm{Z} ,j\in 2\mathbbm{Z}+b\atop
k,l\ge 0, j<0 \, {\rm for}
\, k=l=0}
\Big\{ 1 - e^{2\pi i r / N} \, e^{ 2\pi i ( k \widetilde \sigma + l \widetilde \rho + j \widetilde v) 
}
\Big\}^{ \sum_{s=0}^{N-1}
e^{-2\pi i rs/N}  c^{(0,s)}_b(4kl - j^2)
} \nn , \\ \label{main_twining}
\eea
where for $  r,s\in\mathbbm{Z}, 0\le r,s\le N-1$. The infinite product is such that $k, l$ and $j$ are all integers, $k \ge 0, l \ge 0$. When $k$ and $l$ are  positive integers, $j$ runs over all integers.   When $k$ or $l$ equals 0, $j$ runs over all integers. When both $k$ and $l$ equal $0$, $j$ takes only negative integer values. The coefficients $c^{(r,s)}$ are determined from the functions
\bea
F^{(r,s)}(\tau,z) &\equiv& {1\over N} \mathrm{Tr}_{\mathrm{RR}; g_N^r} \left(  g_N^s
(-1)^{J_L+ J_R}
e^{2\pi i \tau L_0} e^{-2\pi i \bar\tau 
\bar L_0} e^{2\pi i J_L z}\right) \label{Frs_twining} \\
&=&\sum_{b=0}^1\sum_{j\in2\mathbbm{Z}+b, n\in \mathbbm{Z}/N
\atop
4n - j^2\ge -b^2} 
c^{(r,s)}_b(4n -j^2)
e^{2\pi i n\tau + 2\pi i jz}\, .
\eea
In expression \eqref{Frs_twining} 
$\mathrm{Tr}$ denotes trace over all the $g_N^r$ twisted
RR sector 
states in the (4,4) SCFT with K3 as its target
space. $L_0$ and $\bar L_0$ are the left and right-moving Virasoro
generators  and $J_L/2$ and $J_R/2$ are the generators of the
$U(1)_L\times U(1)_R$ subgroup of the $SU(2)_L\times SU(2)_R$
R-symmetry group of this SCFT. For various values of $N$ explicit expressions for $F^{(r,s)}(\tau,z)$ can be found in \cite{Sen:2007qy}. We do not repeat the full expressions here. 

For different $\mathbbm{Z}_{N}$ twinings, our aim is to construct the 4d hair partition functions.  In section \ref{sec:hair_modes} we briefly discuss the hair modes and in section \ref{sec:hair_removal} write the 4d hair partition functions. Finally, using these results we write the twined hair removed black hole partition functions. 

\subsection{Hair modes}
\label{sec:hair_modes}
In \cite{Chakrabarti:2020ugm} a detailed analysis of possible bosonic and fermionic hair modes for the D1-D5 black holes in  CHL models was given. The results from that paper can be readily adapted to twined indices.  

 A hair mode of a black hole is a smooth and normalisable deformation that lives entirely outside the horizon 
 and preserves all the supersymmetries of the black hole. We concentrate on the four-dimensional BMPV black hole.  The four-dimensional BMPV black hole is obtained by placing the five-dimensional BMPV black hole at the center of the Taub-NUT space.  It is most convenient to describe the hair modes as six-dimensional configurations in the supergravity obtained by truncating IIB supergravity on K3. Let $\mathrm{S}^1$ corresponds to $x_5$ and the 
$\widetilde{\mathrm{S}}^1$ to $x_4$.  In \cite{Chakrabarti:2020ugm}, the Gibbons-Hawking coordinates $(r, \theta, \phi, x_4)$ for the Taub-NUT space along with the null coordinates $u=x_5 - t$ and $v=x_5 + t$ are used to describe hair modes.  The hair modes constructed in \cite{Chakrabarti:2020ugm} are all characterised by periodic functions of $v$, i.e., they are all left moving. Three different types of hair modes were constructed for the 4d black holes.

\subsubsection*{Fermionic hair modes}
 The six-dimensional supergravity truncation is a $(2,0)$  theory with 16 supersymmetries. The  black hole solutions preserve 4 of these supersymmetries and hence give rise to 12 fermionic zero modes. Out of these 12 zero modes, four are left moving and 8 are right moving. The 4 left moving modes are elevated to hair modes. These hair modes are characterised by arbitrary functions of $v$ preserving the supersymmetry of the original solution. We construct these modes by solving the linearised equations of motion for the gravitino,
 \be 
 \Gamma^{MNP} D_N \Psi^{\alpha}_P - \bar{H}^{kMNP} \Gamma_N \widehat{\Gamma}^{k}_{\alpha \beta} \Psi^\beta_P = 0, \label{gravitino}
 \ee
 and then showing that the solutions of the linearised equations continue to be solutions of the non-linear equations. 
In equation \eqref{gravitino} $\bar{H}^{kMNP}$ is the self-dual part of the RR form field, $\Gamma$ and $\widehat{\Gamma}$ represent the six-dimensional coordinate space gamma matrices and the Euclidean internal space gamma matrices respectively. For details see \cite{Banerjee:2009uk, Jatkar:2009yd, Chakrabarti:2020ugm}. Using an ansatz for the gravitino $\Psi_P^{\alpha}$, we find that the solution to this equation that corresponds to a hair is given by,
\be
\Psi_v = \psi(r)^{-3/2} h(v) \widetilde{\eta}(\theta, \phi).
\ee
The spinorial properties of the gravitino are completely captured by $\widetilde{\eta}(\theta, \phi)$. The number of independent components of the gravitino turn out to be four. Thus, we have four fermionic hair modes which contribute a factor of
\be
Z^\rom{fermion}_\rom{hair} = \prod_{l = 1}^{\infty} (1 - e^{2 \pi i l \widetilde \rho})^{4} \label{Z_fermion}
\ee
to the hair partition function. We note that $\widetilde \rho$ is the fugacity conjugate to the momentum charge $n$ along the $\text{S}^1$, cf.~\eqref{bilinears}--\eqref{fourier_expansion}.  All these modes are  neutral under the $g_N$ action. These modes also appear for the 5d black holes. In our discussion below it will be convenient to  separate out the $Z^\rom{fermion}_\rom{hair}$ factor.

\subsubsection*{Garfinkle-Vachaspati modes}
 The Garfinkle-Vachaspati transform \cite{Garfinkle:1990jq, Kaloper:1996hr, Mishra:2018bcb, Chakrabarti:2019lfu} is a solution generating technique that adds wave like deformations to a known solution of the bosonic sector equations. Given a metric that possesses a  null, hypersurface orthogonal, Killing vector $k^M$, this technique deforms the original solution as,
\be
G '_{MN} = G_{MN} + e^S \Psi k_M k_N,
\ee
where $S$ is a scalar that is determined from the hypersurface orthogonality condition and $\Psi$ is a scalar (deformation) that satisfies a wave equation with respect to the undeformed metric $G_{MN}$. The four-dimensional BMPV  black hole allows for a smooth deformation in the $G_{vv}$ component alone and is of the form~\cite{Jatkar:2009yd, Chakrabarti:2020ugm}
\be
g_i(v) y^i,
\ee
where $g_i(v)$ are three periodic scalar functions that represent three left moving bosonic hair modes and $y^i$ are the coordinates of the three dimensional transverse space $\mathbb{R}^3$. These hair modes contribute the following to the hair partition function,
\be
Z^\rom{GV}_\rom{hair} = \prod_{l = 1}^{\infty} \frac{1}{(1 - e^{2 \pi i l  \widetilde \rho})^3}.
\ee
All these modes are also neutral under the $g_N$ action.  These modes do not appear for the 5d black holes.

\subsubsection*{Form field hair modes}
In the six-dimensional supergravity truncation, there are $n_t$ tensor multiplets neutral under $g_N$.  In the original black hole solutions (both 4d and 5d) all these tensor multiplets are unexcited, i.e., set to zero. Using the harmonic 2-form $\omega_{TN}$ of the Taub-NUT space these tensor multiplets can be turned on for the 4d black hole as \cite{Banerjee:2009uk, Jatkar:2009yd, Chakrabarti:2020ugm},
\be
\delta H^s_{MNP} = h^s(v) dv \wedge \omega_{TN}, \qquad 1 \leq s \leq n_t. 
\ee
With one deformation $h^s(v)$ for each multiplet $1 \le s \le n_t$, we have $n_t$ such deformations. These deformations are smooth and normalisable and serve as black hole hair. These $n_t$ anti-self-dual (asd)  left moving modes contribute the following to the hair partition function,
\be
Z^\rom{asd}_\rom{hair} = \prod_{l = 1}^{\infty} \frac{1}{(1 - e^{2 \pi i l \widetilde \rho})^{n_t}}.
\ee
 These modes do not appear for the 5d black holes.  The other details of the hair configurations are not essential for the analysis in this paper.  

\subsection{Twining hair removal}

\label{sec:hair_removal}

The total contribution to the hair partition function of modes neutral under $g_N$ is,
\bea
Z_\rom{4d}^\rom{hair}(\widetilde\rho,\widetilde\sigma,\widetilde v) &=&  Z^\rom{fermion}_\rom{hair} \: \: Z^\rom{GV}_\rom{hair} \: \: Z^\rom{asd}_\rom{hair}  \\ 
&=&  Z^\rom{fermion}_\rom{hair}  \: \: \prod_{l=1}^\infty
\left(1 - e^{2\pi i l \widetilde\rho}\right)^{-n_t -3} = \prod_{l=1}^\infty 
\left(1 - e^{2\pi i l \widetilde\rho}\right)^{-n_t +1} \,. \label{Z0}
\eea
For $N \neq 1$ this is not the end of the story. There are additional hair modes. They come from the tensor multiplets \emph{charged} under $g_N$. 
A way to incorporate them in supergravity is to analyse the problem in ten-dimensions \cite{Chakrabarti:2020ugm}. Let us schematically denote $y$ to be the $\text{K3}$ directions and $x$ to be the remaining six dimensions. Then, in ten-dimensions the RR four-form field schematically decomposes as~\cite{hep-th/9506126},
\be
 C_4(x,y) \propto \sum_{\gamma} c^\gamma_{2}(x) \wedge \omega^\gamma(y), \label{C4}
\ee
where  $\omega^\gamma(y)$ are the self-dual and anti-self-dual (asd) harmonic forms spanning the cohomology $H^2(\text{K3}).$ On the elements on this cohomology,  the abelian group of order $N$  generated by $g_N$ acts.   The number of anti-self-dual harmonic forms on K3 with eigenvalue $\exp \left\{-2 \pi i s/N\right\}$ under $g_N$ is denoted $b_s$.  The values of $b_s$ for various twining order $N$ are listed in table \ref{table4}. These numbers are easily obtained from \cite{Chaudhuri:1995dj}; see also table 4 of~\cite{Chakrabarti:2020ugm}. These additional sectors contribute  to the twined hair partition function as, 
\be
 Z^\rom{asd}_{\text{hair}} = \prod_{n=0}^{N-1} Z^{(n)}. 
\ee
where
\be
Z^{(n)}(\widetilde\rho)= \prod_{l =1}^{\infty} (1 - e^{2 \pi i n/N} e^{2 \pi i l \widetilde \rho})^{-(b_n+ 2 \delta_{n,0})}, \qquad \qquad n =0,\ldots, N-1.
\ee

\begin{table}[t!]
\begin{center}
\begin{tabular}{|c||c|c|c|c|c|c|c|c|c|} \hline
$N$ & $b_0$ & $b_1$ & $b_2$ & $b_3$ & $b_4$ & $b_5$ & $b_6$ & $b_7$ \\ \hline \hline
1&19&  &  & & &  & &   \\ \hline
2&11&8&  & & &  & &   \\ \hline
3&7&6&6&  &  &  & &   \\ \hline
4&5 &4&6  &4  &  & &  &   \\ \hline
5&3  &4 &4 &4 &4 &  & &\\ \hline
6&3 &2 &4  &4  &4  &2 &  &   \\ \hline
7&1 &3 &3 &3 &3 &3 &3 & \\ \hline
8&1 &2 &3  &2  &4   &2 &3  &2  \\ \hline
\end{tabular}
\caption{Possible geometric $\mathbbm{Z}_N$ action on K3 cohomology. The numbers $b_s$ denote the number of anti-self-dual $(1,1)$ with eigenvalue $\exp \left\{-2 \pi i s/N\right\}$ under the $g_N$ action. We note that the number of $g_N$ invariant tensor multiplets in  six-dimensional supergravity description is the number of $g_N$ invariant anti-self-dual (1,1) forms $b_0$ plus 2: $n_t = b_0 + 2$. Recall that the plus 2 comes from the self-dual and anti-self-dual decomposition of the RR and NS-NS 2-form fields.}
\label{table4}
\end{center}
\end{table}

For various values of $N$, the hair partition functions are as follows. We use the notation $q = e^{2\pi i \widetilde \rho}$.
For $N=1$,
\begin{align}
Z^{\text{hair}}_{\text{4d}: 1A} = Z^\rom{GV}_\rom{hair} \: \: Z^\rom{fermion}_\rom{hair} \: \: Z^{(0)} =\prod_{n=1}^\infty (1-q^n)^{-20} = \frac{  1 } {\eta(\widetilde \rho)^{24}}  \, e^{2\pi i \widetilde \rho} \, Z^\rom{fermion}_\rom{hair} ,
\end{align}
where it is convenient to separate out the contribution of fermions given in \eqref{Z_fermion}.  For $N=2$,
\begin{align}
Z^{\text{hair}}_{\text{4d}: 2A} &= Z^\rom{GV}_\rom{hair} \: \: Z^\rom{fermion}_\rom{hair} \: \: Z^{(0)} \: \: Z^{(1)} \\ &=   Z^\rom{fermion}_\rom{hair} \prod_{n=1}^\infty (1-q^n)^{-16} (1+q^n)^{-8}\\
&=    Z^\rom{fermion}_\rom{hair}\prod_{n=1}^\infty (1-q^n)^{-8} \Big((1-q^n)(1+q^n)\Big)^{-8}\\
&=  Z^\rom{fermion}_\rom{hair}\prod_{n=1}^\infty (1-q^n)^{-8} (1-q^{2n})^{-8} = \frac1{\eta(\widetilde \rho)^8 \eta(2\widetilde \rho)^8} \, e^{2\pi i \widetilde \rho} \, Z^\rom{fermion}_\rom{hair}.
\end{align}
For $N=3$,
\begin{align}
Z^{\text{hair}}_{\text{4d}: 3A} &= Z^\rom{GV}_\rom{hair} \: \: Z^\rom{fermion}_\rom{hair} \: \:Z^{(0)} \: \:Z^{(1)} \: \: Z^{(2)} \\ &=Z^\rom{fermion}_\rom{hair}  \prod_{n=1}^\infty (1-q^n)^{-12}  (1- e^{2 \pi i/3} q^n)^{-6}  (1- e^{4 \pi i/3} q^n)^{-6} \\
&= Z^\rom{fermion}_\rom{hair}  \prod_{n=1}^\infty (1-q^n)^{-6}  \Big((1-q^n)(1- e^{2 \pi i/3} q^n) (1- e^{4 \pi i/3} q^n)\Big)^{-6} \\
&= Z^\rom{fermion}_\rom{hair}  \prod_{n=1}^\infty  (1-q^n)^{-6} (1-q^{3n})^{-6}  = \frac1{\eta(\widetilde \rho)^6 \eta(3\widetilde \rho)^6} \, e^{2\pi i \widetilde \rho} \, Z^\rom{fermion}_\rom{hair}.
\end{align}
Here, we have made use of the identity $ (1-q^n)(1- \omega q^n) (1- \omega^2 q^n) = (1 - q^{3n})$ with $\omega$ being the third root of unity. We proceed similarly for $N=4,5,6,7,8$. We only write the final answers,
\begin{align}
Z^{\text{hair}}_{\text{4d}: 4B}
&= \frac1{\eta(\widetilde \rho)^4 \eta(2\widetilde \rho)^2 \eta(4\widetilde \rho)^4} \, e^{2\pi i \widetilde \rho} \, Z^\rom{fermion}_\rom{hair}, \\ 
Z^{\text{hair}}_{\text{4d}: 5A}
&= \frac1{\eta(\widetilde \rho)^4 \eta(5 \widetilde \rho)^4} \, e^{2\pi i \widetilde \rho} \, Z^\rom{fermion}_\rom{hair}, \\
Z^{\text{hair}}_{\text{4d}: 6A} 
&= \frac1{\eta(\widetilde \rho)^2 \eta(2\widetilde \rho)^2 \eta(3\widetilde \rho)^2 \eta(6\widetilde \rho)^2 } \, e^{2\pi i \widetilde \rho} \, Z^\rom{fermion}_\rom{hair}, \\
Z^{\text{hair}}_{\text{4d}: 7A} 
&= \frac1{\eta(\widetilde \rho)^3 \eta(7\widetilde \rho)^3} \, e^{2\pi i \widetilde \rho} \, Z^\rom{fermion}_\rom{hair}, \\
Z^{\text{hair}}_{\text{4d}: 8A} 
&= \frac1{\eta(\widetilde \rho)^2 \eta(2 \widetilde \rho) \eta(4\widetilde \rho) \eta(8 \widetilde \rho)^2} \, e^{2\pi i \widetilde \rho} \, Z^\rom{fermion}_\rom{hair}.
\end{align}

The eta-products appearing on the right hand side directly correspond to cycle shapes for certain conjugacy classes of the Mathieu group M$_{24}$. It is now standard in the literature to use these  conjugacy classes to label twined partition functions. Hence the notation $Z^{\text{hair}}_{\text{4d}: 1A}$, $Z^{\text{hair}}_{\text{4d}: 4B}$, etc. Given the results of \cite{Chakrabarti:2020ugm, Chattopadhyaya:2020yzl}, it is clear that the hair removed 4d and 5d partition functions match. We also note that the 4d hair partition functions are closely related to the KK monopole partition functions. In fact, in all cases, the hair partition functions are the KK monopole partition functions with  the additional factor $e^{2\pi i \widetilde \rho} \, Z^\rom{fermion}_\rom{hair}$. Finally, the hair removed twining partition functions are,
\be
\frac{1}{ Z^{\rom{hair}}_\text{4d} }\, \, \frac{1}{ \widetilde \Phi(\widetilde \rho,\widetilde \sigma, \widetilde v)} .
\ee

\setcounter{equation}{0}

\section{Positivity of Fourier coefficients}
\label{sec:positivity_twining}
In this section, we study the positivity of coefficients $S_a$ defined in equation \eqref{linear_comb} for the unorbifolded model. We focus on $N = 2, 3, 4$ twinings. These three cases are sufficiently non-trivial and  capture the essential ideas we wish to convey.   Extension to twinings with higher $N$ is only computationally tedious. In section \ref{sec:positivity_twining_no_hair_removed}, we compute the appropriate Fourier coefficients from the full 4d partition functions.  In a duality frame where there are no hair apart from the fermionic zero modes, the full 4d partition function captures the horizon states. In section \ref{sec:positivity_twining_hair_removed}, 
 we compute the appropriate Fourier coefficients from the hair removed 4d partition functions in the D1-D5 frame. In this frame, the hair removed partition functions are expected\footnote{We cannot rule out the existence of additional hair modes.} to capture the horizon states.

\subsection{Twined partition functions}
\label{sec:positivity_twining_no_hair_removed}

The Fourier coefficients for the Siegel modular form can be  extracted using the contour prescription used in \cite{1008.4209} where we first expand $1/\widetilde{\Phi} (\widetilde \rho, \widetilde \sigma, \widetilde v)$ in powers of $e^{2\pi i \widetilde \rho}$ and  $e^{2\pi i \widetilde \sigma}$ and then expand each term in powers of  $e^{-2\pi i \widetilde v}$. 
The contour together with the following conditions
\be \label{conditions}
Q \cdot P\ge 0, \quad Q \cdot P\le Q^2, \quad 
Q \cdot P\le P^2, \quad Q^2, P^2,
(Q^2 P^2 - (Q \cdot P)^2) > 0\, 
\ee
ensure that the index counts microstates of a finite size single centered black hole.\footnote{The zeros of $\widetilde \Phi$ responsible for wall crossing do not change with twining. Thus the constraints \eqref{conditions} do not change with  twining $N$. }

 For the untwisted model, with no twining our results are presented in table \ref{table_1}. This table is identical to table 1 of \cite{1008.4209}. On a modern computer it takes less than a second to generate entries in table \ref{table_1}.  We only give the results for $Q^2\le P^2$ for all the tables in this section, since
the results are symmetric under $Q^2\leftrightarrow P^2$.
\begin{table}[h!] {\small	
\begin{center} \def\st{\vrule height 3ex width 0ex}
\begin{tabular}{|l|l|l|l|l|l|l|l|l|l|l|} \hline 
\backslashbox{$(Q^2,P^2)$}{$Q \cdot P$}
 &  $-$2 & 0 & 1 & 2 & 3 & 4\st\\[1ex] \hline \hline
(2,2) &   $-$209304 &  {\bf 50064}
&  {\bf 25353} &  648 & 327 & 0 \st\\[1ex] \hline
(2,4) &  $-$2023536  & {\bf 1127472}
&  {\bf 561576} & {\bf 50064} & 8376 & $-$648 \st\\[1ex] \hline
(4,4) & $-$16620544  &  {\bf 32861184} & {\bf 18458000} &  {\bf 3859456}
&  {\bf 561576} & 12800 \st\\[1ex] \hline
(2,6) &  $-$15493728 &  {\bf 16491600}
&  {\bf 8533821} & {\bf 1127472} & 130329 & $-$15600 \st\\[1ex] \hline
(4,6) & $-$53249700 &  {\bf 632078672} & {\bf 392427528}   & 
{\bf 110910300}  &  
{\bf 18458000} 
&  {\bf 1127472} \st\\[1ex] \hline
(6,6) & 2857656828  &  {\bf 16193130552} & {\bf 11232685725}  & 
{\bf 4173501828}
&  {\bf 920577636} & {\bf 110910300}  \st\\[1ex] 
 \hline 
\end{tabular}
\caption{Values of the degeneracy $-B_6$ for the untwisted, untwined model for different values of $Q^2$, $P^2$ and
$Q \cdot P$. The boldfaced entries are for charges that satisfy the constraints \eqref{conditions}. This table is identical to table 1 of \cite{1008.4209}.}
\label{table_1}
\end{center} }
\end{table}

\subsection*{$\mathbbm{Z}_2$ twining}
For the untwisted model, with $\mathbbm{Z}_2$ twining our results for $-B_6^{g_2}$ are presented in table \ref{table_2}.
By taking the sum and the difference of $-B_6$ and $-B_6^{g_2}$, we find  $-S_0$ and $-S_1$ respectively. These values are presented in table \ref{table_2_S0} and \ref{table_2_S1} respectively.

\begin{table}[h!] {\small	
\begin{center} \def\st{\vrule height 3ex width 0ex}
\begin{tabular}{|l|l|l|l|l|l|l|l|l|l|l|} \hline 
\backslashbox{$(Q^2,P^2)$}{$Q \cdot P$}
 &  $-$2 & 0 & 1 & 2 & 3 & 4\st\\[1ex] \hline \hline
(2,2) &   $-$5624 &  $-${\bf 1328}
&  {\bf 505} &  $-$216 & 55 & 0 \st\\[1ex] \hline
(2,4) &  $-$27952  & $-${\bf 7696}
&  {\bf 3128} & $-${\bf 1328} & 488 & $-$104 \st\\[1ex] \hline
(4,4) & $-$138240  &  $-${\bf44544} & {\bf 19120} &  $-${\bf 7168}
&  {\bf 3128} & $-$1280 \st\\[1ex] \hline
(2,6) &  $-$124384 &  $-${\bf 33520}
&  {\bf 14781} &$-${\bf 7696} & 3209 & $-$848 \st\\[1ex] \hline
(4,6) & $-$615780 &  $-${\bf 188528} & {\bf 86232}   & $-${\bf 41316}  &  
{\bf 19120} 
& $-${\bf 7696} \st\\[1ex] \hline
(6,6) & $-$2761380  &  $-${\bf 723144} & {\bf 353853}  & 
$-${\bf 243612}
&  {\bf 126180} & $-${\bf 41316}  \st\\[1ex] 
 \hline 
\end{tabular}
\caption{Values of $-B_6^{g_2}$  for $\mathbbm{Z}_2$ twining for different values of $Q^2$, $P^2$ and
$Q \cdot P$. The boldfaced entries are for charges that satisfy the constraints \eqref{conditions}. }
\label{table_2}
\end{center} }
\end{table}

\begin{table}[h!] {\small	
		\begin{center} \def\st{\vrule height 3ex width 0ex}
			\begin{tabular}{|l|l|l|l|l|l|l|l|l|l|l|} \hline 
				\backslashbox{$(Q^2,P^2)$}{$Q \cdot P$}
				&  $-$2 & 0 & 1 & 2 & 3 & 4\st\\[1ex] \hline \hline
				(2,2) & $-$107464 & {\bf 24368} & {\bf 12929} & 216 & 191 & 0 \st\\[1ex] \hline
				(2,4) & $-$1025744 & {\bf 559888}
				&  {\bf 282352} & {\bf 24368} & 4432 & $-$376 \st\\[1ex] \hline
				(4,4) & $-$8379392 & {\bf 16408320} & {\bf 9238560} & {\bf 1926144}
				& {\bf 282352} & 5760 \st\\[1ex] \hline
				(2,6) & $-$7809056 & {\bf 8229040}
				& {\bf 4274301} & {\bf 559888} & 66769 & $-$8224 \st\\[1ex] \hline
				(4,6) & $-$26932740 & {\bf 315945072} & {\bf 196256880} & {\bf 55434492}  &  
				{\bf 9238560} 
				& {\bf 559888} \st\\[1ex] \hline
				(6,6) & 1427447724 & {\bf 8096203704} & {\bf 5616519789}  & {\bf 2086629108}
				&  {\bf 460351908} & {\bf 55434492}  \st\\[1ex] 
				\hline 
			\end{tabular} 
\caption{Values of $-S_0$  with $\mathbbm{Z}_2$ twining for different values of $Q^2$, $P^2$ and
$Q \cdot P$. The boldfaced entries are for charges that satisfy the constraints \eqref{conditions}.  Note that all the boldfaced entries are positive integers. }
\label{table_2_S0}
\end{center} }
\end{table}

\begin{table}[h!] {\small	
		\begin{center} \def\st{\vrule height 3ex width 0ex}
			\begin{tabular}{|l|l|l|l|l|l|l|l|l|l|l|} \hline 
				\backslashbox{$(Q^2,P^2)$}{$Q \cdot P$}
				&  $-$2 & 0 & 1 & 2 & 3 & 4\st\\[1ex] \hline \hline
				(2,2) & $-$101840 & {\bf 25696} & {\bf 12424} & 432 & 136 & 0 \st\\[1ex] \hline
				(2,4) & $-$997792 & {\bf 567584}
				&  {\bf 279224} & {\bf 25696} & 3944 & $-$272 \st\\[1ex] \hline
				(4,4) & $-$8241152 & {\bf 16452864} & {\bf 9219440} & {\bf 1933312}
				& {\bf 279224} & 7040 \st\\[1ex] \hline
				(2,6) & $-$7684672 & {\bf 8262560}
				& {\bf 4259520} & {\bf 567584} & 63560 & $-$7376 \st\\[1ex] \hline
				(4,6) & $-$26316960 & {\bf 316133600} & {\bf 196170648} & {\bf 55475808}  &  
				{\bf 9219440} 
				& {\bf 567584} \st\\[1ex] \hline
				(6,6) & 1430209104 & {\bf 8096926848} & {\bf 5616165936}  & {\bf 2086872720}
				&  {\bf 460225728} & {\bf 55475808}  \st\\[1ex] 
				\hline 
			\end{tabular} 
\caption{Values of $-S_1$  with $\mathbbm{Z}_2$ twining for different values of $Q^2$, $P^2$ and
$Q \cdot P$. The boldfaced entries are for charges that satisfy the constraints \eqref{conditions}.  Note that all the boldfaced entries are positive integers. We note that the number of states that contribute to the total degeneracy with eigenvalue $+1$ (table \ref{table_2_S0}) and $-1$ (this table) are approximately the same. }
\label{table_2_S1}
\end{center} }
\end{table}

\subsubsection*{$\mathbbm{Z}_3$ twining}
For the untwisted model, with $\mathbbm{Z}_3$ twining our results for $-B_6^{g_3}$ are presented in table \ref{table_3}. We note that the values for $-B_6^{g^2_3}$ are the same as $-B_6^{g_3}$. Thus, 
using $-B_6$ and $-B_6^{g_3}$ we can easily compute $-S_0$, $-S_1$, and $-S_2$. We find $-S_1$ and $-S_2$ are identical. The values for $-S_0$, $-S_1$ are presented in table \ref{table_3_S0} and \ref{table_3_S1} respectively.

\begin{table}[h!] {\small	
\begin{center} \def\st{\vrule height 3ex width 0ex}
\begin{tabular}{|l|l|l|l|l|l|l|l|l|l|l|} \hline 
\backslashbox{$(Q^2,P^2)$}{$Q \cdot P$}
 &  $-$2 & 0 & 1 & 2 & 3 & 4\st\\[1ex] \hline \hline
(2,2) &   $-$1566 &  $-${\bf 588}
&  {\bf 297} &  $-$108 & 30 & 0 \st\\[1ex] \hline
(2,4) &  $-$6204  & $-${\bf 2442}
&  {\bf 1272} & $-${\bf 588} & 204 & $-$54 \st\\[1ex] \hline
(4,4) & $-$24328  &  $-${\bf 9696} & {\bf 5390} &  $-${\bf 2696}
&  {\bf 1272} & $-$448 \st\\[1ex] \hline
(2,6) &  $-$21396 &  $-${\bf 8964}
&  {\bf 4998} &$-${\bf 2442} & 1026 & $-$336 \st\\[1ex] \hline
(4,6) & $-$83964 &  $-${\bf 35446} & {\bf 20256}   & $-${\bf 10956}  &  
{\bf 5390} 
& $-${\bf 2442} \st\\[1ex] \hline
(6,6) & $-$288510  &  $-${\bf 127332} & {\bf 76209}  & 
$-${\bf 42108}
&  {\bf 22545} & $-${\bf 10956}  \st\\[1ex] 
 \hline 
\end{tabular}
\caption{Values of $-B_6^{g_3} \equiv -B_6^{g^2_3}$  for $\mathbbm{Z}_3$ twining for different values of $Q^2$, $P^2$ and
$Q \cdot P$. The boldfaced entries are for charges that satisfy the constraints \eqref{conditions}. }
\label{table_3}
\end{center} }
\end{table}

\begin{table}[h!] {\small	
		\begin{center} \def\st{\vrule height 3ex width 0ex}
			\begin{tabular}{|l|l|l|l|l|l|l|l|l|l|l|} \hline 
				\backslashbox{$(Q^2,P^2)$}{$Q \cdot P$}
				&  $-$2 & 0 & 1 & 2 & 3 & 4\st\\[1ex] \hline \hline
				(2,2) & $-$70812 &  {\bf 16296}
				&  {\bf 8649} & 144 & 129 & 0 \st\\[1ex] \hline
				(2,4) &  $-$678648  & {\bf 374196}
				&  {\bf 188040} & {\bf 16296} & 2928 & $-$252 \st\\[1ex] \hline
				(4,4) & $-$5556400  &  {\bf 10947264} & {\bf 6156260} &  {\bf 1284688}
				&  {\bf 188040} & 3968 \st\\[1ex] \hline
				(2,6) & $-$5178840 &  {\bf 5491224}
				&  {\bf 2847939} & {\bf 374196} & 44127 & $-$5424 \st\\[1ex] \hline
				(4,6) & $-$17805876 &  {\bf 210669260} & {\bf 130822680}   &
				{\bf 36962796}  &  
				{\bf 6156260} 
				& {\bf 374196} \st\\[1ex] \hline
				(6,6) & 952359936 &  {\bf 5397625296} & {\bf 3744279381}  & 
				{\bf 1391139204}
				&  {\bf 306874242} & {\bf 36962796}  \st\\[1ex] 
				\hline 
			\end{tabular}
\caption{Values of $-S_0$  with $\mathbbm{Z}_3$ twining for different values of $Q^2$, $P^2$ and
$Q \cdot P$. The boldfaced entries are for charges that satisfy the constraints \eqref{conditions}.  Note that all the boldfaced entries are positive integers. }
\label{table_3_S0}
\end{center} }
\end{table}

\begin{table}[h!] {\small	
		\begin{center} \def\st{\vrule height 3ex width 0ex}
			\begin{tabular}{|l|l|l|l|l|l|l|l|l|l|l|} \hline 
				\backslashbox{$(Q^2,P^2)$}{$Q \cdot P$}
				&  $-$2 & 0 & 1 & 2 & 3 & 4\st\\[1ex] \hline \hline
				(2,2) & $-$69246 &  {\bf 16884}
				&  {\bf 8352} & 252 & 99 & 0 \st\\[1ex] \hline
				(2,4) & $-$672444 & {\bf 376638}
				&  {\bf 186768} & {\bf 16884} & 2724 &  $-$198 \st\\[1ex] \hline
				(4,4) & $-$5532072 &  {\bf 10956960} & {\bf 6150870} &  {\bf 1287384}
				&  {\bf 186768} & 4416 \st\\[1ex] \hline
				(2,6) & $-$5157444 &  {\bf 5500188}
				&  {\bf 2842941} & {\bf 376638} & 43101 & $-$5088 \st\\[1ex] \hline
				(4,6) & $-$17721912 &  {\bf 210704706} & {\bf 130802424}   &
				{\bf 36973752}  &  
				{\bf 6150870} 
				& {\bf 376638} \st\\[1ex] \hline
				(6,6) & 952648446 &  {\bf 5397752628} & {\bf 3744203172}  & 
				{\bf 1391181312}
				&  {\bf 306851697} & {\bf 36973752}  \st\\[1ex] 
				\hline 
			\end{tabular} 
\caption{Values of $-S_1 \equiv -S_2$  with $\mathbbm{Z}_3$ twining for different values of $Q^2$, $P^2$ and
$Q \cdot P$. The boldfaced entries are for charges that satisfy the constraints \eqref{conditions}.  Note that all the boldfaced entries are positive integers. We also note that the number of states that contribute to the degeneracy with eigenvalue $+1$ (table \ref{table_3_S0}) and eigenvalues $e^{2\pi i/3}$ or $e^{4\pi i/3}$ (this table) are approximately the same. }
\label{table_3_S1}
\end{center} }
\end{table}

\subsubsection*{$\mathbbm{Z}_4$ twining}
For the untwisted model, with $\mathbbm{Z}_4$ twining our results for $-B_6^{g_4}$ are presented in table \ref{table_4}.
We note that the values for $-B_6^{g^2_4}$ are the same as $-B_6^{g_2}$, already presented in table \ref{table_2}. 
Furthermore, we note that the values for $-B_6^{g^3_4}$ are the same as $-B_6^{g_4}$.
Thus, we can easily compute $-S_a$, $ 0 \le a \le 3$. We find $-S_1$ and $-S_3$ to be identical. The values for $-S_0$, $-S_1$, , $-S_2$ are presented in table \ref{table_4_S0}, table \ref{table_4_S1}, and table \ref{table_4_S2} respectively.

\begin{table}[h!] {\small	
\begin{center} \def\st{\vrule height 3ex width 0ex}
\begin{tabular}{|l|l|l|l|l|l|l|l|l|l|l|} \hline 
\backslashbox{$(Q^2,P^2)$}{$Q \cdot P$}
 &  $-$2 & 0 & 1 & 2 & 3 & 4\st\\[1ex] \hline \hline
(2,2) &   $-$560 &  $-${\bf 224}
&  {\bf 117} &  $-$48 & 19 & 0 \st\\[1ex] \hline
(2,4) &  $-$1760  & $-${\bf 768}
&  {\bf 444} & $-${\bf 224} & 84 & $-$32 \st\\[1ex] \hline
(4,4) & $-$5504  &  $-${\bf 2624} & {\bf 1608} &  $-${\bf 896}
&  {\bf 444} & $-$160 \st\\[1ex] \hline
(2,6) &  $-$5312 &  $-${\bf 2400}
&  {\bf 1437} &$-${\bf 768} & 373 & $-$136 \st\\[1ex] \hline
(4,6) & $-$16660 &  $-${\bf 8128} & {\bf 5148}   & $-${\bf 3028}  &  
{\bf 1608} 
& $-${\bf 768} \st\\[1ex] \hline
(6,6) & $-$50156  &  $-${\bf 24712} & {\bf 16117}  & 
$-${\bf 9828}
&  {\bf 5652} & $-${\bf 3028}  \st\\[1ex] 
 \hline 
\end{tabular}
\caption{Values of $-B_6^{g_4} \equiv -B_6^{g^3_4}$  for $\mathbbm{Z}_4$ twining for different values of $Q^2$, $P^2$ and $Q \cdot P$. The boldfaced entries are for charges that satisfy the constraints \eqref{conditions}.}
\label{table_4}
\end{center} }
\end{table}

\begin{table}[h!] {\small	
		\begin{center} \def\st{\vrule height 3ex width 0ex}
			\begin{tabular}{|l|l|l|l|l|l|l|l|l|l|l|} \hline 
				\backslashbox{$(Q^2,P^2)$}{$Q \cdot P$}
				&  $-$2 & 0 & 1 & 2 & 3 & 4\st\\[1ex] \hline \hline
				(2,2) & $-$54012 &  {\bf 12072}
				&  {\bf 6523} & 84 & 105 & 0 \st\\[1ex] \hline
				(2,4) & $-$513752 & {\bf 279560}
				&  {\bf 141398} & {\bf12072} & 2258 & $-$204 \st\\[1ex] \hline
				(4,4) & $-$4192448 &  {\bf 8202848} & {\bf 4620084} &  {\bf 962624}
				&  {\bf 141398} & 2800 \st\\[1ex] \hline
				(2,6) & $-$3907184 &  {\bf 4113320}
				&  {\bf 2137869} & {\bf 279560} & 33571 &  $-$4180\st\\[1ex] \hline
				(4,6) & $-$13474700 &  {\bf 157968472} & {\bf 98131014}   &
				{\bf 27715732}  &  
				{\bf 4620084} 
				& {\bf 279560} \st\\[1ex] \hline
				(6,6) & 713698784 &  {\bf 4048089496} & {\bf 2808267953}  & 
				{\bf 1043309640}
				&  {\bf 230178780} & {\bf27715732}  \st\\[1ex] 
				\hline 
			\end{tabular}
\caption{Values of $-S_0$  with $\mathbbm{Z}_4$ twining for different values of $Q^2$, $P^2$ and
$Q \cdot P$. The boldfaced entries are for charges that satisfy the constraints \eqref{conditions}.  Note that all the boldfaced entries are positive integers. }
\label{table_4_S0}
\end{center} }
\end{table}

\begin{table}[h!] {\small	
\begin{center} \def\st{\vrule height 3ex width 0ex}
\begin{tabular}{|l|l|l|l|l|l|l|l|l|l|l|} \hline 
\backslashbox{$(Q^2,P^2)$}{$Q \cdot P$}
 &  $-$2 & 0 & 1 & 2 & 3 & 4\st\\[1ex] \hline \hline
(2,2) & $-$50920 & 12848 & 6212 & 216 & 68 & 0 \st\\[1ex] \hline
(2,4) & $-$498896 & {\bf 283792} &  {\bf 139612} & {\bf 12848} & 1972 & $-$136 \st\\[1ex] \hline
(4,4) & $-$4120576  & {\bf 8226432} & {\bf 4609720} & {\bf 966656}
&  {\bf 139612} & 3520 \st\\[1ex] \hline
(2,6) &  $-$3842336 & {\bf 4131280} &  {\bf 2129760} & {\bf 283792} & 31780 & $-$3688 \st\\[1ex] \hline
(4,6) & $-$13158480 &  {\bf 158066800} & {\bf 98085324}   & {\bf 27737904}  & {\bf 4609720} & {\bf 283792} \st\\[1ex] \hline
(6,6) & 715104552 & {\bf 4048463424} & {\bf 2808082968}  & {\bf 1043436360} & {\bf 230112864} & {\bf 27737904}  \st\\[1ex] 
 \hline 
\end{tabular}
\caption{Values of $-S_1$ and $-S_3$ with $\mathbbm{Z}_4$ twining for different values of $Q^2$, $P^2$ and
$Q \cdot P$. The boldfaced entries are for charges that satisfy the constraints \eqref{conditions}.  Note that all the boldfaced entries are positive integers. }
\label{table_4_S1}
\end{center} }
\end{table}

\begin{table}[h!] {\small	
\begin{center} \def\st{\vrule height 3ex width 0ex}
\begin{tabular}{|l|l|l|l|l|l|l|l|l|l|l|} \hline 
\backslashbox{$(Q^2,P^2)$}{$Q \cdot P$}
 &  $-$2 & 0 & 1 & 2 & 3 & 4\st\\[1ex] \hline \hline
(2,2) &  $-$53452 & 12296 & 6406 & 132 & 86 & 0 \st\\[1ex] \hline
(2,4) &  $-$511992 & {\bf 280328} & {\bf 140954} & {\bf 12296} & 2174 & $-$172 \st\\[1ex] \hline
(4,4) & $-$4186944  & {\bf 8205472} & {\bf 4618476} & {\bf 963520} &  {\bf 140954} & 2960 \st\\[1ex] \hline
(2,6) & $-$3901872 & {\bf 4115720} &  {\bf 2136432} & {\bf 280328} & 33198 & $-$4044 \st\\[1ex] \hline
(4,6) & $-$13458040 & {\bf 157976600} & {\bf 98125866}   & {\bf 27718760}  & {\bf 4618476} 
& {\bf 280328} \st\\[1ex] \hline
(6,6) & 713748940 & {\bf 4048114208} & {\bf 2808251836}  & {\bf 1043319468}
&  {\bf 230173128} & {\bf 27718760}  \st\\[1ex] 
 \hline 
\end{tabular}
\caption{Values of $-S_2$ for $\mathbbm{Z}_4$ twining for different values of $Q^2$, $P^2$ and
$Q \cdot P$. The boldfaced entries are for charges that satisfy the constraints \eqref{conditions}.  Note that all the boldfaced entries are positive integers. We also note that the number of states that contribute to the degeneracy with eigenvalue $+1$ (table \ref{table_4_S0}), eigenvalues $e^{\pi i/2}$ or $e^{3\pi i/2}$ (table \ref{table_4_S1}), and  eigenvalue $-1$ (this table) are approximately the same.}
\label{table_4_S2}
\end{center} }
\end{table}

\subsubsection*{Comment about the implementation}
The above tables were constructed using Sen's formula \eqref{main_twining}, and we verified that they match with the Cl\'ery-Gritsenko formula \cite{Gritsenko:2008, Govindarajan:2011em, Govindarajan:2020owu}. Implementation of Cl\'ery-Gritsenko formula is in fact easier in \verb+Mathematica+ and computation time is shorter, especially for the higher values of $N$. We do not present  details about the Cl\'ery-Gritsenko formula here.

\subsection{Hair removed twined partition functions}

\label{sec:positivity_twining_hair_removed}

The hair removed twining partition functions are,
\be
Z_
\text{horizon} = \frac{1}{ Z^{\rom{hair}}_\text{4d} }\, \, \frac{1}{ \widetilde \Phi(\widetilde \rho,\widetilde \sigma, \widetilde v)} . 
\ee
 The Fourier coefficients $-B_6^g {}_{:\text{horizon}}$ can be extracted from $Z_\text{horizon}$  using the contour prescription mentioned above. These coefficients are computed in tables \ref{table_N1_hair-removed} (no twining), \ref{table_N2_hair-removed} ($\mathbbm{Z}_2$ twining), \ref{table_N3_hair-removed} ($\mathbbm{Z}_3$ twining), \ref{table_N4_hair-removed} ($\mathbbm{Z}_4$ twining).  Table \ref{table_N1_hair-removed} (no twining) is identical to table 4 of \cite{Chattopadhyaya:2020yzl}. In this section we do not present tables for degeneracy of horizon states with definite $g_N$ eigenvalues. These numbers can be easily constructed by taking the linear combinations of tables given. We have checked that positivity property holds as expected.

 \begin{table}[h!] {\small	
\begin{center} \def\st{\vrule height 3ex width 0ex}
\begin{tabular}{|l|l|l|l|l|l|l|l|l|l|l|} \hline 
\backslashbox{$(Q^2,P^2)$}{$Q \cdot P$}
 &  $-$2 & 0 & 1 & 2 & 3 & 4\st\\[1ex] \hline \hline
(2,2) &   $-$7464 &  {\bf 28944}
&  {\bf 13863} &  1608 & 327 & 0 \st\\[1ex] \hline
(2,4) &  $-$17176  & {\bf 761312}
&  {\bf 406296} & {\bf 72424} & 6936 & $-$648 \st\\[1ex] \hline
(4,4) & 2409376  &  {\bf 12980224} & {\bf 8595680} &  {\bf 2665376}
&  {\bf 406296} & 25760 \st\\[1ex] \hline
(2,6) &  704952 &  {\bf 12324920}
&  {\bf 6995541} & {\bf 1423152} & 96619 & $-$13680 \st\\[1ex] \hline
(4,6) & 83729820 &  {\bf 333276712} & {\bf 235492308} & {\bf 85781820}   & 
{\bf 16141380}  &  
{\bf 1423152} 
\st\\[1ex] \hline
(6,6) & 2153280528  &  {\bf 6227822652} & {\bf 4771720755}  & 
{\bf 2158667028}
&  {\bf 572268361} & {\bf 85781820}  \st\\[1ex] 
 \hline 
\end{tabular}
\caption{Values of coefficients (the analog of $-B_6$) for the hair-removed partition function without twining  for different values of $Q^2$, $P^2$ and
$Q \cdot P$. The boldfaced entries are for charges that satisfy the constraints \eqref{conditions}. This table is identical to table 4 of \cite{Chattopadhyaya:2020yzl}. }
\label{table_N1_hair-removed}
\end{center} }
\end{table}

\begin{table}[h!] {\small	
\begin{center} \def\st{\vrule height 3ex width 0ex}
\begin{tabular}{|l|l|l|l|l|l|l|l|l|l|l|} \hline 
\backslashbox{$(Q^2,P^2)$}{$Q \cdot P$}
 &  $-$2 & 0 & 1 & 2 & 3 & 4\st\\[1ex] \hline \hline
(2,2) &   $-$1608 &  $-${\bf 368}
&  {\bf 199} &  $-$152 & 55 & 0 \st\\[1ex] \hline
(2,4) &  $-$7960  & $-${\bf 1952}
&  {\bf 1032} & $-${\bf 792} & 392 & $-$104 \st\\[1ex] \hline
(4,4) & $-$21728  &  $-${\bf 10240} & {\bf 4640} &  $-${\bf 1248}
&  {\bf 1032} & $-$864 \st\\[1ex] \hline
(2,6) &  $-$35528 &  $-${\bf 6664}
&  {\bf 3701} &$-${\bf 4112} & 2411 & $-$720 \st\\[1ex] \hline
(4,6) & $-$96484 &  $-${\bf 39448} & {\bf 18132}   & $-$
{\bf 6724}  &  
{\bf 5332} 
& $-${\bf 4112} \st\\[1ex] \hline
(6,6) & $-$301680  &  $-${\bf 4996} & {\bf 21523}  & 
$-${\bf 66060}
&  {\bf 38121} & $-${\bf 6724}  \st\\[1ex] 
 \hline 
\end{tabular}
\caption{Values of coefficients (the analog of $-B_6^{g_2}$) for the hair-removed partition function with $\mathbbm{Z}_2$ twining for different values of $Q^2$, $P^2$ and
$Q \cdot P$. The boldfaced entries are for charges that satisfy the constraints \eqref{conditions}.}
\label{table_N2_hair-removed}
\end{center} }
\end{table}

\begin{table}[h!] {\small	
\begin{center} \def\st{\vrule height 3ex width 0ex}
\begin{tabular}{|l|l|l|l|l|l|l|l|l|l|l|} \hline 
\backslashbox{$(Q^2,P^2)$}{$Q \cdot P$}
 &  $-$2 & 0 & 1 & 2 & 3 & 4\st\\[1ex] \hline \hline
(2,2) &   $-$840 &  $-${\bf 360}
&  {\bf 210} &  $-$84 & 30 & 0 \st\\[1ex] \hline
(2,4) &  $-$3352  & $-${\bf 1492}
&  {\bf 846} & $-${\bf 440} & 168 & $-$54 \st\\[1ex] \hline
(4,4) & $-$9752  &  $-${\bf 4148} & {\bf 2570} &  $-${\bf 1432}
&  {\bf 846} & $-$340 \st\\[1ex] \hline
(2,6) &  $-$11574 &  $-${\bf 5404}
&  {\bf 3261} &$-${\bf 1746} & 805 & $-$288 \st\\[1ex] \hline
(4,6) & $-$33720 &  $-${\bf 14996} & {\bf 9108}   & $-$
{\bf 5640}  &  
{\bf 3210} 
& $-${\bf 1746} \st\\[1ex] \hline
(6,6) & $-$90432  &  $-${\bf 43536} & {\bf 28440}  & 
$-${\bf 16722}
&  {\bf 10372} & $-${\bf 5640}  \st\\[1ex] 
 \hline 
\end{tabular}
\caption{Values of coefficients (the analog of $-B_6^{g_3}$) for the hair-removed partition function with $\mathbbm{Z}_3$ twining for different values of $Q^2$, $P^2$ and
$Q \cdot P$. The boldfaced entries are for charges that satisfy the constraints \eqref{conditions}.}
\label{table_N3_hair-removed}
\end{center} }
\end{table}

\begin{table}[h!] {\small	
\begin{center} \def\st{\vrule height 3ex width 0ex}
\begin{tabular}{|l|l|l|l|l|l|l|l|l|l|l|} \hline 
\backslashbox{$(Q^2,P^2)$}{$Q \cdot P$}
 &  $-$2 & 0 & 1 & 2 & 3 & 4\st\\[1ex] \hline \hline
(2,2) &   $-$496 &  $-${\bf 208}
&  {\bf 111} &  $-$48 & 19 & 0 \st\\[1ex] \hline
(2,4) &  $-$1560  & $-${\bf 704}
&  {\bf 420} & $-${\bf 216} & 84 & $-$32 \st\\[1ex] \hline
(4,4) & $-$4672  &  $-${\bf 2304} & {\bf 1440} &  $-${\bf 832}
&  {\bf 420} & $-$160 \st\\[1ex] \hline
(2,6) &  $-$4712 &  $-${\bf 2200}
&  {\bf 1341} &$-${\bf 736} & 363 & $-$136 \st\\[1ex] \hline
(4,6) & $-$14116 &  $-${\bf 7080} & {\bf 4596}   & $-$
{\bf 2756}  &  
{\bf 1512} 
& $-${\bf 736} \st\\[1ex] \hline
(6,6) & $-$38032  &  $-${\bf 19412} & {\bf 13003}  & 
$-${\bf 8212}
&  {\bf 4881} & $-${\bf 2756}  \st\\[1ex] 
 \hline 
\end{tabular}
\caption{Values of coefficients (the analog of $-B_6^{g_4}$) for the hair-removed partition function with $\mathbbm{Z}_4$ twining for different values of $Q^2$, $P^2$ and
$Q \cdot P$. The boldfaced entries are for charges that satisfy the constraints \eqref{conditions}.}
\label{table_N4_hair-removed}
\end{center} }
\end{table}

\setcounter{equation}{0}

\section{Twisted-twining hair removal}
\label{sec:twisted_twining}   

Having discussed a class of twined partition functions of K3 compactification, we now turn  to the twined partition functions of  the CHL models. The CHL orbifolds are often called twisted models. Accordingly, we call the twined partition functions for these models twisted-twining partition functions.

We consider type IIB string theory compactified on K3$~\times~ \widetilde{\mathrm{S}}^1~\times~\mathrm{S}^1$ and mod out this theory by a $\mathbbm{Z}_{M}$ symmetry generated by $1/M$ shift along the S$^1$ and an order $M$ transformation $g_M$ on K3. We take the radius of S$^1$ to be $M$ and the radius of $ \widetilde{\mathrm{S}}^1$ to be 1.  Momentum along the  S$^1$ circle is quantised in units of $1/M$. 
We consider the D1-D5 system as described in detail in \cite{Sen:2007qy, Chakrabarti:2020ugm}. For such a set-up T-duality invariant charge bilinears are
\begin{align}
Q^2 &= 2n/M, & P^2 &= 2 (Q_1 - 1), & Q \cdot P = J.
\end{align}

In a subspace of the moduli space of this compactification, we identify $g_N$ to be the generator of a specific geometric $\mathbbm{Z}_{N}$ symmetry of K3 that preserves all the covariantly constant spinors of K3 and leaves invariant some 2-cycles of K3.  A complete list of possible symmetries of this type can be found in \cite{Chaudhuri:1995dj, hep-th/9508154}. 
From these papers we learn that there are a total of 7 cases to consider, they are listed in table \ref{table_models}.

For this set-up, in the region of the moduli space where the type IIB string coupling is small, the result for the twined index $B^{g_N}_{6}$ is,
\be 
-B^{g_N}_{6}  =  (-1)^{Q\cdot P+1}\,
g\left({M\over 2} Q^2 , {1\over 2\, M}\, P^2,
Q\cdot P\right),
\ee
where $g(l,k,j)$ are
the coefficients of Fourier expansion of the function
$1/ \widetilde \Phi(\widetilde \rho,\widetilde \sigma, \widetilde v)$.   The function $\widetilde\Phi(\widetilde \rho,\widetilde \sigma,\widetilde v)$ is a modular form
of a subgroup of Sp$(2,\ZZZ)$, given by~\cite{1002.3857},
\bea 
\widetilde \Phi(\widetilde \rho, \widetilde \sigma, \widetilde v) &=&  e^{2\pi i (\widetilde \alpha \widetilde \rho
+\widetilde \gamma \widetilde \sigma + \widetilde \beta \widetilde v)} \,
\prod_{b=0}^1 \prod_{r=0}^{N-1}\prod_{r'=0}^{M-1}
\prod_{k\in\zzz+{r'\over M}, l\in\zzz, j\in 2\ZZZ+b\atop
k,l\ge 0, j<0 \, for\, k=l=0}
\left[ 1 - e^{2\pi i r/N} \, e^{2\pi i (k \widetilde \sigma +l \widetilde \rho + j \widetilde v)}
\right]^a\nonumber \\
a &\equiv& \sum_{s=0}^{N-1} \sum_{s'=0}^{M-1} e^{-2 
\pi i (s'l / M +
rs/N)} c_b^{(0,s;r',s')}(4kl - j^2)\, . \label{twisted_twining_main}
\eea
Here too, the infinite product is to be understood as before. The point of distinction is in the fact that when $r' \neq 0$, $k$  takes fractional values. The coefficients $c_b^{(r,s;r',s')}$ are defined via the 
equation:
\bea \label{Frsrpsp2}
F^{(r,s;r',s')}(\tau, z) &=& {1\over MN} \mathrm{Tr}_{\mathrm{RR};g_M^{r'} g_N^r} \left (g_M^{s'} g_N^{s} 
(-1)^{J_L+J_R} e^{2\pi i (\tau L_0-\bar\tau \bar L_0)}
e^{2\pi i J_L z} \right)\, \\
&=&\sum_{b=0}^1 \sum_{j\in 2\ZZZ+b, n\in \ZZZ/MN} c_b^{(r,s;r',s')}
(4n - j^2) e^{2\pi i (n\tau + j z)}. \label{Frsrpsp}
\eea
In equation \eqref{Frsrpsp2}, the trace is over all the $g_M^{r'} g_N^{r}$ twisted
RR sector states in the (4,4) superconformal  CFT
with target space K3.  The coefficients 
$\widetilde\alpha$, $\widetilde\beta$, $\widetilde\gamma$ are given by
\bea
\widetilde\alpha &=& {1\over 24M} Q_{0,0} -{1\over 2M}
\sum_{s'=1}^{M-1} Q_{0,s'} {e^{-2\pi i s'/M}
\over (1- e^{-2\pi i s'/M}
)^2}\, , \label{alpha}  \\
\widetilde\beta &=& 1,  \\
\widetilde\gamma &=& {1\over 24 M}\, \chi(\text{K3}) = {1\over 24 M}\, Q_{0,0}\, ,
\eea
where
\be \label{def_qrs}
Q_{r',s'} = M N \left( c_0^{(0,0;r',s')}(0)
+ 2  c_1^{(0,0;r',s')}(-1) \right),
\ee
and where the Euler number of  $\text{K3}$, $\chi(\text{K3}) =24$.

 \subsection{Hair modes for the $\ZZZ_M~\times~\ZZZ_N$ models}

Now we wish to discuss the hair modes relevant for the twisted-twining partition functions. As  in section \ref{sec:twining}, there are three types of hair modes to consider:  fermionic, Garfinkle-Vachaspati, and modes from the tensor-multiplet sectors.  The fermionic and  Garfinkle-Vachaspati modes are all neutral under the $\ZZZ_M~\times~\ZZZ_N$ action.  It is only the tensor-multiplet sector that requires a detailed consideration. 

\begin{table}
\begin{center}
\begin{tabular}{|c|c|c|c|c|c|c|} \hline 
$\mathbbm{Z}_2 \times \mathbbm{Z}_2$ &
$\mathbbm{Z}_3 \times \mathbbm{Z}_3$ &
$\mathbbm{Z}_2 \times \mathbbm{Z}_4 $ &
$\mathbbm{Z}_4 \times \mathbbm{Z}_2 $ &
$\mathbbm{Z}_4 \times \mathbbm{Z}_4$ &
$\mathbbm{Z}_2 \times \mathbbm{Z}_6 $ &
$\mathbbm{Z}_6 \times \mathbbm{Z}_2 $ \\
 \hline 
\end{tabular}
\caption{ The list of $\ZZZ_M~\times~\ZZZ_N$ symmetries that can be geometrically realised in an appropriate subspace of the moduli space of K3.} \label{table_models}
\end{center} 
\end{table}


At this stage, it is instructive to  quickly recall the hair partition functions for the twisted cases, with no twining  \cite{Chakrabarti:2020ugm, Chattopadhyaya:2020yzl}.  For these cases, in addition to the untwisted hair partition function \eqref{Z0}, there are contributions from the twisted sectors. 
To incorporate the contributions from the twisted sectors, it is convenient to analyse the problem in ten-dimensions.
In ten-dimensions, the RR four-form field schematically decomposes as \eqref{C4},
\be
 C_4(x,y) \propto \sum_{\gamma} c^\gamma_{2}(x) \wedge \omega^\gamma(y), \label{C4_2}
\ee
 where $\omega^\gamma(y)$ are the self-dual and anti-self-dual harmonic forms spanning the cohomology $H^2(\text{K3}).$ On these harmonic forms,  the abelian orbifold group of order $M$  acts.    Since $\omega^\gamma(y)$ are not all $g_M$ invariant, the fields $c^\gamma_{2}(x)$ pick up the opposite  phases under the CHL orbifold action.    The combined effect ensures that the ten-dimensional $ C_4(x,y) $ is $g_M$ invariant.

Such modes contribute to the hair partition functions. In order to account for their contributions, we must know the number of tensor-multiplets transforming with eigenvalue  
$e^{2 \pi i m/M}$ for $0 \le m \le M-1$ under $g_M$. This data is given in table \ref{table4}. The contribution to the 4d hair partition function due to the such modes  is of the form~\cite{Chakrabarti:2020ugm},
\begin{equation}
  Z_{(m)} = \prod_{l=1}^{\infty} (1 - e^{-2 \pi i m \widetilde \rho} e^{2 \pi i M l \widetilde \rho})^{-(b_m + 2 \delta_{m,0})} \;\;\; ,\;\;m =0,1,\cdots M-1,
\end{equation}
with the full $4$d hair partition function given by the product,
\begin{equation}
    Z^{\text{hair}}_{\text{4d}} = Z_{\text{hair}}^\rom{GV} \: Z_{\text{hair}}^\rom{fermion} \prod_{m=0}^{M-1} Z_{(m)},
\end{equation}
where
\be
Z_{\text{hair}}^\rom{fermion} = \prod_{l=1}^\infty \, (1-e^{2\pi i M l \widetilde \rho})^4,
\ee
and
\be
Z_{\text{hair}}^\rom{GV} = \prod_{l=1}^\infty \, (1-e^{2\pi i M l \widetilde \rho})^{-3}.
\ee
Further details can be found in \cite{Chakrabarti:2020ugm}.

From the above discussion, it is clear that  we need to know how many of the tensor multiplets are charged under $\ZZZ_M~\times~\ZZZ_N$ with eigenvalues  $e^{\frac{2 \pi i m}{M}}~\mbox{and}~e^{\frac{2 \pi i n}{N}}$ respectively for $ 0 \le m \le M -1 $ and $0 \le n \le N -1$.  To find these numbers, we need to
diagonalise the action of  
$\ZZZ_M~\times~\ZZZ_N$ on the 19  anti-self-dual (1,1) forms of K3.  Let $p_m^n$ be the 
number of tensor-multiplets with eigenvalues $e^{\frac{2 \pi i m}{M}}$  with respect to $\ZZZ_M$ action and $e^{\frac{2 \pi i n}{N}}$ with respect to $\ZZZ_N$ action.  From \cite{Chaudhuri:1995dj} we can work out these decompositions. The 19 anti-self-dual (1,1) forms decompose as: 
\paragraph{$\mathbbm{Z}_2~\times~\mathbbm{Z}_2$} :
$
19 =7_0^0 + 4_0^1 + 4_1^0 + 4_1^1.
$
\paragraph{$\mathbbm{Z}_3~\times~\mathbbm{Z}_3$} :
$
19 
= 3_0^0 + 2_0^1 + 2_0^2 + 2_1^0 + 2_1^1 + 2_1^2 + 2_2^0 + 2_2^1 + 2_2^2.
$
\paragraph{$\mathbbm{Z}_2~\times~\mathbbm{Z}_4$} :
$
19 = 3^0_0 + 2^1_0 + 4^2_0 + 2^3_0 + 2^0_1 + 2^1_1   + 2^2_1 + 2^3_1.
$
 \paragraph{$\mathbbm{Z}_4~\times~\mathbbm{Z}_2$} :
$
19 =  3^0_0 + 2_0^1 + 2_1^0 + 2_1^1 + 4_2^0 + 2_2^1 + 2_3^0 + 2_3^1 .
$
\paragraph{$\mathbbm{Z}_4~\times~\mathbbm{Z}_4$} :
 $
 19 = 1_0^0 + 1_0^1  + 2_0^2 + 1_0^3 + 1_1^0 + 1_1^1 + 1_1^2 + 1_1^3 + 2_2^0 + 1_2^1 + 2_2^2+ 1_2^3 + 1_3^0 + 1_3^1 + 1_3^2 + 1_3^3.
$

 \paragraph{$\mathbbm{Z}_2~\times~\mathbbm{Z}_6$} :
$ 19 = 1_0^0 + 1_0^1 + 3_0^2 + 2_0^3 +3_0^4+1_0^5 +2_1^0+1_1^1 +1_1^2 +2_1^3 +1_1^4+1_1^5 .$

 \paragraph{$\mathbbm{Z}_6~\times~\mathbbm{Z}_2$} :
$ 19 = 1_0^0 + 2^1_0  + 1^0_1  + 1^1_1 + 3^0_2 + 1^1_2 + 2^0_3 + 2^1_3 + 3^0_4 + 1^1_4 +1^0_5 + 1^1_5 .$

\subsection{Hair removed twisted-twining partition functions}

\label{sec:hair_removed_twisted_twining}
The following product of various factors give the twisted-twining 4d hair partition functions,
\be
 Z^{\rom{hair}}_\text{4d} = Z^\rom{GV}_\rom{hair}  \, Z^\rom{fermion}_\rom{hair}  \prod_{n=0}^{N-1}\prod_{m=0}^{M-1}  Z_m^n,
\ee
where
 \be
 Z_m^n = \prod_{l=1}^\infty \left( 1 -  e^{\frac{2 \pi i n}{N}} q^{M l-m} \right)^{-p_m^n - 2 \delta_m^0  \delta_0^n},  \qquad 0 \le m \le M-1, \qquad 0 \le n \le N-1.
 \ee
 Here too, we use the standard notation $q = e^{2 \pi i \widetilde \rho}$. For the $\mathbbm{Z}_2~\times~\mathbbm{Z}_2$ model, it takes the form
\bea
 Z^{\rom{hair}}_\text{4d} &=&  Z^\rom{GV}_\rom{hair}  \: \: Z^\rom{fermion}_\rom{hair}  \: \: Z_0^0 \: \: Z_0^1 \: \: Z_1^0 \: \: Z_1^1 \\
 &=&  Z^\rom{fermion}_\rom{hair}  \prod_{l=1}^\infty 
  \left( 1 - q^{2l}\right)^{-12}
  \left( 1 + q^{2 l} \right)^{-4}
 \left( 1 - q^{2l -1}\right)^{-4}
  \left( 1 + q^{2 l-1} \right)^{-4} \\
  &=&   Z^\rom{fermion}_\rom{hair} \prod_{l=1}^\infty 
  \left( 1 - q^{2l}\right)^{-8}
  \left( 1 + q^{l} \right)^{-4}
 \left( 1 - q^{l}\right)^{-4} \\
   &=&   Z^\rom{fermion}_\rom{hair} \prod_{l=1}^\infty 
  \left( 1 - q^{2l}\right)^{-8}
  \left( 1 - q^{2l} \right)^{-4} \\
   &=&  Z^\rom{fermion}_\rom{hair}  \prod_{l=1}^\infty 
  \left( 1 - q^{2l}\right)^{-12} \\
  &=&  Z^\rom{fermion}_\rom{hair} e^{2 \pi i \widetilde \rho}  \frac{1}{ \eta ( 2 \widetilde \rho)^{12}} .
\eea

As before, the hair partition function is closely related to the KK monopole partition function for this model.  Apart from the factor $Z^\rom{fermion}_\rom{hair} e^{2 \pi i \widetilde \rho}$ it \textit{is} the KK monopole partition function. There are multiple ways of confirming this. We show it here using the expression for  the KK monopole partition function as given in \cite{1002.3857},
 \be 
Z_\text{KK} = e^{-2\pi i\widetilde\alpha \widetilde \rho}\,
\prod_{r=0}^{N-1} \prod_{l=1}^\infty \left(1 - e^{2\pi i r/N}
e^{2\pi i l \widetilde \rho}\right)^{-\sum_{s=0}^{N-1}\sum_{s'=0}^{M-1} 
e^{-2\pi i r s/N} 
e^{-2\pi i ls'/M} \left(c_0^{(0,s;0,s')}(0) + 2 c_1^{(0,s;
0,s')}(-1)
\right)}\, , \nn \\
\ee
where $c_1^{(0,s;0,s')}(-1) = 2/MN$ and the coefficients $c_0^{(0,s;0,s')}(0)$ are to be found from the functions $F^{(0,s; 0,s')}$. The variable $\widetilde\alpha$ was introduced in \eqref{alpha}.

The $F^{(0,s; 0,s')}$ functions for the $\mathbbm{Z}_2~\times~\mathbbm{Z}_2$ model are as follows. The function $F^{(0,0;0,0)} $ is 
  \be
F^{(0,0;0,0)}(\tau, z) = 2 A(\tau,z) , \label{eq:F0000}
\ee
where $A(\tau,z)$  is written in terms of the Jacobi theta functions $\vartheta_i$ as,
\be
A(\tau,z) = \Bigg[ \frac{\vartheta_2(\tau,z)^2}{\vartheta_2(\tau,0)^2} + \frac{\vartheta_3(\tau,z)^2}{\vartheta_3(\tau,0)^2} + \frac{\vartheta_4(\tau,z)^2}{\vartheta_4(\tau,0)^2} \Bigg]. \label{eq:A}
\ee
For completeness, we recall that the four most common Jacobi theta functions are defined by
\be
\theta \bigg[ \begin{matrix} a \\ b \end{matrix} \bigg] (\tau, z) = \sum_{l \in \mathbbm{Z}} \hat q^{\half ( l + \frac{a}{2} )^2}  \hat r^{ ( l + \frac{a}{2} )} e^{i \pi l b}, 
\ee
where $a \cdot b = (0,1)$  mod 2. In this notation, $\vartheta_1 {(\tau, z)} \equiv \theta \bigg[ \begin{matrix} 1 \\ 1 \end{matrix} \bigg] (\tau, z)$, $\vartheta_2 {(\tau, z)} \equiv \theta \bigg[ \begin{matrix} 1 \\ 0 \end{matrix} \bigg] (\tau, z)$, $\vartheta_3 {(\tau, z)} \equiv \theta \bigg[ \begin{matrix} 0 \\ 0 \end{matrix} \bigg] (\tau, z)$ and $\vartheta_4 {(\tau, z)} \equiv \theta \bigg[ \begin{matrix} 0 \\ 1 \end{matrix} \bigg] (\tau, z)$ and where $\hat q = e^{2 \pi i \tau}$ and $\hat r = e^{2 \pi i z}$. 
Furthermore, $2F^{(r,s; 0,0)} =  F^{(r,s)} = 2 F^{(0,0;r',s')}$, where the functions $F^{(r,s)}$ were introduced in \eqref{Frs_twining}.  These functions are well known for several models \cite{hep-th/0602254}. We have,
\be
F^{(0,0;0,1)}(\tau, z) = F^{(0,1; 0,0)}(\tau, z)= \tfrac12 F^{(0,1)}(\tau, z) = \frac{2}{3} A(\tau, z) - \frac{1}{3} B(\tau, z) E_2(\tau). 
\ee
Here, $B(\tau, z) = \eta(\tau)^{-6} \vartheta_1(\tau,z)^2$ and the Eisenstein series $E_N(\tau)$ is defined as
\be
E_N(\tau) = \frac{12 i}{\pi (N-1)} \partial_{\tau}[\ln \eta{(\tau)} - \ln \eta{(N \tau)}] = 1 + \frac{24}{N-1} \sum_{\substack{n_1, n_2 \geq 1 \\ n_1 \neq 0 \text{mod} N}} n_1 e^{2 \pi i n_1 n_2 \tau}. \label{eq:EN}
\ee
The action of the group element $g_M g_N$ on the anti-self-dual (1,1) forms allows us to identify $F^{(0,1;0,1)}$ as an EOT Jacobi form. We find 
 that $F^{(0,1;0,1)} = F^{(0,0;0,1)}$. For the remaining  functions, we rely on the SL$(2,\ZZZ)$ transformations acting on $F^{(r,s;r',s')}$:
\be \label{eq:Frsrs_SL2}
F^{(r,s;r',s')} \left(\frac{a\tau+b}{c\tau+d}, \frac{z}{c\tau+d}\right) = \exp\left(2\pi i\frac{cz^2}{c\tau+d}\right) F^{(cs+ar, ds+br; cs'+ar',ds'+br')} (\tau,z)\ .
\ee

 From these expressions we find that $4c_0^{(0,0;0,0)}(0) = 20$ and  $4c_0^{(0,s;0,s')}(0) = 4$ for other values of $s, s'$.  We find $\widetilde \alpha = 1$ from \eqref{alpha}. Inserting these values in the above formula, after some calculation we  get
\begin{align}
Z_\text{KK} (\widetilde{\rho})  = \frac1{\eta(2\widetilde \rho)^{12}}.
\end{align}

All the other cases are dealt with similarly. Of course, the calculations become more and more tedious. A useful and often-times simpler way to figure out the functions $F^{(0,s; 0, s')}$ is as follows. We note from \eqref{Frsrpsp}  that the trace is over untwisted RR sector with the insertion $g_M^{s'} g_N^{s}$. Since we know the action of $g_M$ and $g_N$ separately on the 24 dimensional cohomology of K3 we can easily work out the trace of $g_M^{s'} g_N^{s}$ over this 24 dimensional representation. The 24-dimensional trace and the order of group element $g_M^{s'} g_N^{s}$ uniquely fixes the $F^{(0,s; 0, s')} (\tau, z) $ as an Eguchi-Ooguri-Tachikawa (EOT) Jacobi form \cite{Eguchi:2010ej}. 

In all cases we find,
\be
 Z^{\rom{hair}}_\text{4d}   =  Z^\rom{fermion}_\rom{hair}  \: e^{2 \pi i \widetilde \rho}   \: Z_\text{KK},
\ee
where $ Z_\text{KK}$ for various models are as follows:
\begin{align}
& \mathbbm{Z}_3~\times~\mathbbm{Z}_3 && {Z_\text{KK}} = \frac1{\eta(3\widetilde \rho)^{8}} & \\
& \mathbbm{Z}_2~\times~\mathbbm{Z}_4 && {Z_\text{KK}} = \frac1{\eta(2\widetilde \rho)^4 \eta(4\widetilde \rho)^4}\ & \\
& \mathbbm{Z}_4~\times~\mathbbm{Z}_2 && {Z_\text{KK}} = \frac1{\eta(2\widetilde \rho)^4 \eta(4\widetilde \rho)^4}\ &  \\
& \mathbbm{Z}_4~\times~\mathbbm{Z}_4 && Z_\text{KK}= \frac1{\eta(4\widetilde \rho)^{6}} & \\
& \mathbbm{Z}_2~\times~\mathbbm{Z}_6 && Z_\text{KK}=  \frac1{\eta(2\widetilde \rho)^3 \eta(6\widetilde \rho)^3}\ & \\
& \mathbbm{Z}_6~\times~\mathbbm{Z}_2 && Z_\text{KK}=  \frac1{\eta(2\widetilde \rho)^3 \eta(6\widetilde \rho)^3}\ &
\end{align}
Finally, the hair removed twisted-twining partition functions are,
\be
\frac{1}{ Z^{\rom{hair}}_\text{4d} }\, \, \frac{1}{ \widetilde \Phi(\widetilde \rho,\widetilde \sigma, \widetilde v)} .
\ee
 
 \setcounter{equation}{0}

\section{Fourier coefficients for the $\ZZZ_2~\times~\ZZZ_2$ model and positivity checks}

\label{sec:positivity_twisted_twining}

Product representation \eqref{twisted_twining_main} gives Siegel modular forms describing twisted-twining partition functions for the $\mathbbm{Z}_M~\times~\mathbbm{Z}_N$ models. Although this formula has a clear physical interpretation, it is cumbersome to work with when it comes to extracting Fourier coefficients. The complexity lies in knowing the functions $F^{(r,s;r',s')}$ whose Fourier coefficients $c^{(r,s;r',s')}$ enter as exponents of the various factors.  Fortunately, there are other representations available for some models, including a  product representation involving genus-two theta functions \cite{1006.3472} and  a product formula using weak Jacobi forms \cite{Gritsenko:1996tm,Gritsenko:2008}. These alternative representations allow for an easier extraction of the Fourier coefficients. In this section, we briefly describe these three different representations for the $\mathbbm{Z}_2~\times~\mathbbm{Z}_2$ model and extract the Fourier coefficients and check the positivity properties. As an important consistency check, we confirm that all these formulae give the same answers. We also extract the Fourier coefficients for the hair removed twisted-twining partition function in the D1-D5 duality frame and check the positivity properties.

\subsection{Sen's product representation}

To work with the product representation \eqref{twisted_twining_main}, we need eight functions $F^{(0,s;r',s')}$ from which we get our $c^{(0,s;r',s')}$.
We already noted in equation \eqref{eq:F0000} that
\be
F^{(0,0;0,0)}(\tau,z)= 2 A(\tau,z),
\ee
where $A(\tau,z)$  is written in terms of the Jacobi theta functions $\vartheta_i$ \eqref{eq:A}. 
As noted in section \ref{sec:hair_removed_twisted_twining}, 
some of the other $F^{(r,s;r',s')}$ can be obtained using the property: $F^{(r,s; 0,0)} = \frac{1}{2} F^{(r,s)} = F^{(0,0;r',s')}$, where the functions $F^{(r,s)}$ were introduced in \eqref{Frs_twining}. 
We have,
\be
F^{(0,0;0,1)}(\tau, z) = F^{(0,1; 0,0)}(\tau, z) = \frac12 F^{(0,1)}(\tau, z) = \frac{2}{3} A(\tau, z) - \frac{1}{3} B(\tau, z) E_2(\tau),
\ee 
where the notations $B(\tau, z)$ and $E_2(\tau)$ are introduced around equation \eqref{eq:EN}. 
Similarly, $F^{(0,0;1,0)}$ and $F^{(0,0;1,1)}$ can be obtained from the function $F^{(r,rk)}$ with $r =1$ and $k= 0$ and $1$, respectively,
\bea
&& F^{(0,0;1,0)}(\tau, z) =  \frac12 F^{(1,0)}(\tau, z) = \frac{2}{3} A(\tau, z) + \frac{1}{6} B(\tau, z) E_2\left(\frac{\tau}{2}\right),\\
&& F^{(0,0;1,1)}(\tau, z) =  \frac12 F^{(1,1)}(\tau, z) = \frac{2}{3} A(\tau, z) + \frac{1}{6} B(\tau, z) E_2\left( \frac{\tau+1}{2} \right).
\eea 
The action of the group element $g_M g_N$ on the anti-self-dual (1,1) forms allows us to identify $F^{(0,1;0,1)}$ as an EOT Jacobi form. We find that $F^{(0,1;0,1)} = F^{(0,0;0,1)}$.
For the remaining two functions, we rely on the $\text{SL}(2,\mathbbm{Z})$ transformations \eqref{eq:Frsrs_SL2} of $F^{(r,s;r',s')}$. 
This gives 
$F^{(0,1;1,1)}(\tau,z) = (F^{(0,1;1,0)}| ST^{-1}ST^{-1}S) (\tau,z)$, where $S: \tau \to -\frac{1}{\tau}$ and $T: \tau \to \tau + 1$.  Finally, from  \cite{Gaberdiel:2012gf, 1312.0622} we  infer that $F^{(0,1;1,0)} (\tau,z) = 0$. With these functions at hand, we have an explicit expression of the Siegel modular form, which can be programmed in \verb+Mathematica+.

\subsection{Product of genus two theta functions representation}

\label{sec:genus_two_theta}

Another representation of the $\mathbbm{Z}_2~\times~\mathbbm{Z}_2$ twisted-twining partition function  is via a product of genus-two theta functions \cite{1006.3472}. 
To introduce genus-two theta functions, we start by recalling that the Siegel upper half-space of genus $2$,  $\mathbbm{H}_2$,  is the set of $2 \times 2$ symmetric matrices over the complex numbers whose imaginary part is positive definite, i.e.,
\be
\mathbf{Z} = \begin{pmatrix} \tau  & z \\ z  & \sigma \end{pmatrix},
\ee 
with imaginary part of $\mathbf{Z}$ positive definite. 
Genus-two theta functions on $\mathbbm{H}_2$ are defined  as,
\be
\theta \bigg[ \begin{matrix} \mathbf a \\ \mathbf b \end{matrix} \bigg] (\mathbf{Z}) = \sum_{l_1, \, l_2 \in \mathbbm{Z}} \hat q^{\half ( l_1 + \frac{a_1}{2} )^2}  \hat r^{ ( l_1 + \frac{a_1}{2} ) ( l_2 + \frac{a_2}{2} )} \hat s^{\half ( l_2 + \frac{a_2}{2} )^2} e^{i \pi (l_1 b_1 + l_2 b_2)},
\ee
where $\mathbf{a} = \bigg( \begin{matrix} a_1 \\ a_2 \end{matrix} \bigg)$, $ \mathbf{b} = \bigg( \begin{matrix}  b_1 \\ b_2 \end{matrix} \bigg)$, and where $\hat q = e^{2 \pi i \tau}$, $\hat r = e^{2 \pi i z}$, and $\hat s = e^{2 \pi i \sigma}$. 

By taking appropriate products of these  functions, we can construct a class of Siegel modular forms. The twisted-twining partition function for the $\mathbbm{Z}_2~\times~\mathbbm{Z}_2$ model can be written as the inverse of the Siegel modular form,
\be
\widetilde{\Phi} (\mathbf{Z}) = \Bigg( \frac{1}{16} \theta \Bigg[ \begin{smallmatrix} 0 \\ 1 \\ 0 \\ 0 \end{smallmatrix}\Bigg] (\mathbf{Z}) \:  \theta \Bigg[ \begin{smallmatrix} 1 \\ 1 \\ 0 \\ 0 \end{smallmatrix} \Bigg] (\mathbf{Z}) \: \theta \Bigg[ \begin{smallmatrix} 0 \\ 1 \\ 1 \\ 0 \end{smallmatrix} \Bigg] (\mathbf{Z}) \: \theta \Bigg[ \begin{smallmatrix} 1 \\ 1 \\ 1 \\ 1 \end{smallmatrix} \Bigg] (\mathbf{Z}) \Bigg)^2.
\ee

 These expressions allow us to extract the desired Fourier coefficients in \verb+Mathematica+. The identification with Sen's product representation notation is,
\be \label{dictionary}
\mathbf{Z} = \begin{pmatrix} \tau  & z \\ z  & \sigma \end{pmatrix}  = \begin{pmatrix} \widetilde \sigma    & \widetilde v \\ \widetilde v  & \widetilde \rho \end{pmatrix}.
\ee 
For some other twisted-twining partition functions too, genus two theta function product representation is known. We will discuss one more example  in section \ref{sec:other_models}.

\subsection{Borcherds product representation}

The product representation described in \cite[Theorem 2.1]{Gritsenko:1996tm} gives another convenient representation for the twisted-twining  partition function for the $\mathbbm{Z}_2~\times~\mathbbm{Z}_2$ model.
Let $\varphi$ be a weak Jacobi form of weight 0 and index $t$
with integral coefficients,
\be
\varphi(\tau ,z)= \sum_{n,l} c(n,l) \hat q^n \hat r^l,
\ee
where $\hat q = e^{2 \pi i \tau}$ and $\hat r = e^{2 \pi i z}$. Define
\be
A = \frac{1}{24}\sum_l c(0,l)~,\qquad  B = \frac{1}{2}\sum_{l>0} l c(0,l)~, \qquad C = \frac{1}{4}\sum_l l^2 c(0,l)\ .
\ee
Then, the Jacobi form $\varphi$ gives a Siegel modular form via the product\footnote{We can call this formula the ``Borcherds lift''  or the ``exponential-lift'' following \cite[section 2.1]{Gritsenko:1996tm}. In the following sections, we  refer to this formula as the Borcherds product formula or the Borcherds lift.}  
\be\label{explift}
\widetilde \Phi (\widetilde \rho, \widetilde \sigma, \widetilde v) = q^A r^B s^C \prod_{
		(n,l,m)>0} (1-q^{tn} r^l s^{m})^{c(nm,l)}\ ,
\ee
where now we again denote $q = e^{2 \pi i \widetilde \rho}, ~r = e^{2 \pi i \widetilde v}$, and $s = e^{2 \pi i \widetilde \sigma}$  with $n, l$ and $m$ all integers. The notation  $(n,l,m)>0$ means   ($n > 0$, $m> 0$, $l \in \ZZZ$) $\cup$ ($n = 0, m > 0$, $l \in \ZZZ$) $\cup$ ($n > 0, m = 0$, $l \in \ZZZ$) $\cup$ ($n = m =0, l < 0$).  The Jacobi forms $\varphi$ for various twisted-twining partition functions of our interest were given in \cite{1006.3472}. For a more recent and more complete discussion see \cite{Govindarajan:2019ezd}.

The index 2 weight 0 Jacobi form that gives the twisted-twining partition function for the $\mathbbm{Z}_2~\times~\mathbbm{Z}_2$ model is,
\be
\varphi = 2 \varphi_1^{(3)} = 4 (f_2^2 f_3^2 + f_3^2 f_4^2 + f_4^2 f_2^2),
\ee
where the notation $\varphi_1^{(3)}$ comes from \cite{Cheng:2013wca}, and where
\be
f_i = \frac{\vartheta_i(\tau, z)}{\vartheta_i(\tau, 0)}, \qquad i \in {2,3,4}.
\ee
With the coefficients $c(nm,l)$ in hand,  the infinite products \eqref{explift} takes the form
\be
\widetilde \Phi (\widetilde \rho, \widetilde \sigma, \widetilde v) = q r s^{1/2} \prod_{(n,l,m) > 0} (1 - q^{2n} r^l s^m)^{c(nm,l)}.
\ee 
We expand this infinite product to get the Fourier coefficients. We also note that while computing coefficients $A, B$ and $C$, the contribution to $c(0,l)$  only comes from $l = 0, \pm 1$.

\subsection{Fourier coefficients}

\begin{table}[h!] {\small
\begin{center}\def\st{\vrule height 3ex width 0ex}
\begin{tabular}{|l|l|l|l|l|l|l|l|l|l|l|} \hline 
\backslashbox{$(Q^2,P^2)$}{$Q \cdot P$}
&  $-$2 
& 0 & 1 & 2 & 3 & 4\st\\[1ex] \hline \hline
(1,2) & $-$5410   &  {\bf 2164}
&  {\bf 360} &  $-$2 & 0 & 0 \st\\[1ex] \hline
(1,4) & $-$26464    & {\bf 18944}
&  {\bf 4352} & {160} & 0 & 0 \st\\[1ex] \hline
(2,4) & $-$124160  &  {\bf  198144} & {\bf  67008} &  {\bf 6912}
&  {64} &  0 \st\\[1ex] \hline
(1,6) & $-$114524  &  {\bf  125860}
&  {\bf  36024} & {\bf 2164} & 52  & 0  \st\\[1ex] \hline
(2,6) & $-$473088  &  {\bf 1580672} & {\bf 671744}   & 
{\bf 101376}  &  
{\bf 4352} 
&  {$-$16} \st\\[1ex] \hline
(3,6) &  $-$779104 &  {\bf 15219528} & {\bf 7997655}  & 
{\bf 1738664}
&  {\bf 149226} & {\bf 2164}  \st\\[1ex] \hline
 \hline 
\end{tabular}
\caption{Values of $-B_6$ for the $\ZZZ_2$ CHL orbifold model 
for different values of $Q^2$, $P^2$ and
$Q \cdot P$. The boldfaced entries are for charges
that satisfy the constraints (\ref{constraints_2}). This table is identical to table 2 of \cite{1008.4209}.  } \label{table_Z2_Z2_1}
\end{center} }
\end{table} 

In any of the above three representations of the same function, we can extract  Fourier coefficients using the contour prescription discussed in section \ref{sec:positivity_twining_no_hair_removed}. 
Since the zeros of $\widetilde \Phi$ responsible for wall crossing do not change with twining, the constraints on charges
that ensure that the index counts single centre black hole microstates 
 are the same as the $\mathbbm{Z}_2$ CHL model \cite{1002.3857}. These constraints are \cite{1008.4209}:
\be \label{constraints_2}
Q \cdot P\ge 0, \quad Q \cdot P\le 2\, Q^2, \quad 
Q \cdot P\le P^2, \quad
3 \, Q \cdot P \le 2 \, Q^2 + P^2,
\quad Q^2, P^2,
\{Q^2 P^2 - (Q \cdot P)^2\} > 0.
\ee
Our results for the Fourier coefficients are summarised in tables \ref{table_Z2_Z2_1} and \ref{table_Z2_Z2_2}.  We only give the results for $2 Q^2 \le P^2$, as  the indices have a symmetry under $P^2 \leftrightarrow 2 Q^2$.  For the hair removed partition functions our results are summarised in tables \ref{table_Z2_Z1_hair_removed} and \ref{table_Z2_Z2_hair_removed}.

\begin{table}[h!] {\small
\begin{center}\def\st{\vrule height 3ex width 0ex}
\begin{tabular}{|l|l|l|l|l|l|l|l|l|l|l|} \hline 
\backslashbox{$(Q^2,P^2)$}{$Q \cdot P$}
&  $-$2 
& 0 & 1 & 2 & 3 & 4\st\\[1ex] \hline \hline
(1,2) & $-$290   &  $-${\bf 12}
& $-$ {\bf 8} &  $-$2 & 0 & 0 \st\\[1ex] \hline
(1,4) & 0   & {\bf 0}
&  {\bf 0} & {0} & 0 & 0 \st\\[1ex] \hline
(2,4) & 0  &  {\bf  0} & {\bf  0} &  {\bf 0}
&  {0} &  0 \st\\[1ex] \hline
(1,6) & $-$2172  &  $-${\bf  12}
&  $-${\bf  120} & $-${\bf 12} & 12  & 0  \st\\[1ex] \hline
(2,6) & 0 &  {\bf 0} & {\bf 0}   & 
{\bf 0}  &  
{\bf 0} 
&  0 \st\\[1ex] \hline
(3,6) &  $-$16512 &  {\bf 2376} & $-${\bf 2217}  & 
$-${\bf 312}
&  {\bf 378} & $-${\bf 12}  \st\\[1ex] \hline
 \hline 
\end{tabular}
\caption{Values of $-B_6^{g_2}$ for the  $\ZZZ_2$-twined partition function of the $\ZZZ_2$ CHL model  
for different values of $Q^2$, $P^2$ and
$Q \cdot P$. The boldfaced entries are for charges
that satisfy the constraints (\ref{constraints_2}). All the boldfaced entries are very small compared to the corresponding entries in table \ref{table_Z2_Z2_1}. Thus, it is clear that the coefficients $-S_0$ and $-S_1$ satisfy the expected positivity property. } \label{table_Z2_Z2_2}
\end{center} }
\end{table}

\begin{table}[h!] {\small
\begin{center}\def\st{\vrule height 3ex width 0ex}
\begin{tabular}{|l|l|l|l|l|l|l|l|l|l|l|} \hline 
\backslashbox{$(Q^2,P^2)$}{$Q \cdot P$}
&  $-$2 
& 0 & 1 & 2 & 3 & 4\st\\[1ex] \hline \hline
(1,2) & $-$418   &  {\bf 852}
&  {\bf 296} &  $-$2 & 0 & 0 \st\\[1ex] \hline
(1,4) & $-$1888    & {\bf 9472}
&  {\bf 3840} & {160} & 0 & 0 \st\\[1ex] \hline
(2,4) & 5632  &  {\bf  64512} & {\bf  33216} &  {\bf 5632}
&  {64} &  0 \st\\[1ex] \hline
(1,6) & $-$6684  &  {\bf  73508}
&  {\bf  32680} & {\bf 2292} & 52  & 0  \st\\[1ex] \hline
(2,6) & 83808  &  {\bf 671840} & {\bf 390432}   & 
{\bf 83808}  &  
{\bf 3936} 
&  {$-$16} \st\\[1ex] \hline
(3,6) &  930352 &  {\bf 4806056} & {\bf 3213211}  & 
{\bf 961768}
&  {\bf 115242} & {\bf 2292}  \st\\[1ex] \hline
 \hline 
\end{tabular}
\caption{Values of coefficients (the analog of $-B_6$) for the hair removed $\ZZZ_2$ CHL orbifold partition function  
for different values of $Q^2$, $P^2$ and
$Q \cdot P$. The boldfaced entries are for charges
that satisfy the constraints (\ref{constraints_2}). This table is not identical to table 5 of \cite{Chattopadhyaya:2020yzl}; although they are computing the same quantities. See footnote \ref{footnote:AJ}. } \label{table_Z2_Z1_hair_removed}
\end{center} }
\end{table}

\begin{table}[h!] {\small
\begin{center}\def\st{\vrule height 3ex width 0ex}
\begin{tabular}{|l|l|l|l|l|l|l|l|l|l|l|} \hline 
\backslashbox{$(Q^2,P^2)$}{$Q \cdot P$}
&  $-$2 
& 0 & 1 & 2 & 3 & 4\st\\[1ex] \hline \hline
(1,2) & $-$98   & {\bf 4}
& $-${\bf 8} &  $-$2 & 0 & 0 \st\\[1ex] \hline
(1,4) & 0   & {\bf 0}
&  {\bf 0} & {0} & 0 & 0 \st\\[1ex] \hline
(2,4) & 0  &  {\bf  0} & {\bf  0} &  {\bf 0}
&  {0} &  0 \st\\[1ex] \hline
(1,6) & $-$732  &  {\bf  180}
&  $-${\bf  144} & $-${\bf 12} & 12  & 0  \st\\[1ex] \hline
(2,6) & 0 &  {\bf 0} & {\bf 0}   & 
{\bf 0}  &  
{\bf 0} 
&  0 \st\\[1ex] \hline
(3,6) &  $-$2736 &  {\bf 1992} & $-${\bf 1197}  & 
$-${\bf 216}
&  {\bf 282} & $-${\bf 12}  \st\\[1ex] \hline
 \hline 
\end{tabular}
\caption{Values of coefficients (the analog of $-B_6^{g_2}$) for the  hair-removed $\ZZZ_2$-twined partition function of the $\ZZZ_2$ CHL model  
for different values of $Q^2$, $P^2$ and
$Q \cdot P$. The boldfaced entries are for charges
that satisfy the constraints (\ref{constraints_2}). All the boldfaced entries are very small compared to the corresponding entries in table \ref{table_Z2_Z1_hair_removed}. Thus, the hair removed analog of coefficients $-S_0$ and $-S_1$ satisfy the expected positivity property. } \label{table_Z2_Z2_hair_removed}
\end{center} }
\end{table}

\setcounter{equation}{0}

\section{Positivity checks for  other models}
\label{sec:other_models}

There are at least two other cases that can be dealt with as a straightforward extension of the techniques and results obtained so far. These are $\ZZZ_2~\times~\ZZZ_4$ and $\ZZZ_3~\times~\ZZZ_3$. We check the positivity properties for the indices for these models in sections \ref{sec:Z2Z4} and \ref{sec:Z3Z3} respectively. As noted earlier, the Sen's product representation for the twisted-twining partition functions  for these models  is fairly cumbersome. For example, for the $\ZZZ_3~\times~\ZZZ_3$ model, we would need $F^{(0,s;r',s')}$ functions for $s,r',s' \in \{0,1, 2\}$ (27 functions in total) in order to extract the Fourier coefficients. Fortunately, there are other representations available for these models, including a  product representation involving genus-two theta functions \cite{1006.3472}. These alternative representations allow for an easier extraction of the Fourier coefficients, which is what we use.

There are two other cases that can also be dealt with using our techniques straightforwardly, namely $\ZZZ_4~\times~\ZZZ_2$ and $\ZZZ_4~\times~\ZZZ_4$. However, to the best of our knowledge, the precise conditions on the charge vectors to describe the single centered black holes have not been worked out.  Thus, although we can easily obtain the Fourier coefficients, the interpretation as indices of single center black holes is not fully clear.  For this reason, we do not present results for these other models.

\subsection{$\ZZZ_2~\times~\ZZZ_4$}

\label{sec:Z2Z4}
The twisted-twining partition function for $\ZZZ_2~\times~\ZZZ_4$ model is $\widetilde{\Phi} (\mathbf{Z})^{-1}$ where \cite{1006.3472}
\be
\widetilde{\Phi} (\mathbf{Z}) =  \bigg( \frac{1}{4} \theta \bigg[ \begin{smallmatrix} 0 \\ 1 \\ 1 \\ 0 \end{smallmatrix} \bigg]  (\mathbf{Z}) \: \theta \bigg[ \begin{smallmatrix} 1 \\ 1 \\ 1 \\ 1 \end{smallmatrix} \bigg]  (\mathbf{Z}) \bigg)^2. 
\ee 
This product can be programmed in \verb+Mathematica+, where we recall the dictionary  \eqref{dictionary} between the genus two theta function notation and our previous notation. We find Fourier coefficients as listed in table \ref{table_Z2_Z4}.  Together with previously obtained tables \ref{table_Z2_Z2_1} and \ref{table_Z2_Z2_2} for the $\ZZZ_2$ CHL model, we can easily check that the positivity properties are satisfied  for $-S_0, -S_1, -S_2, -S_3$. Although it is expected on physical grounds, it is quite remarkable that various things conspire to give $-S_0, -S_1, -S_2, -S_3$ positive integers. 

\begin{table}[h!] {\small
\begin{center}\def\st{\vrule height 3ex width 0ex}
\begin{tabular}{|l|l|l|l|l|l|l|l|l|l|l|} \hline 
\backslashbox{$(Q^2,P^2)$}{$Q \cdot P$}
&  $-$2 
& 0 & 1 & 2 & 3 & 4\st\\[1ex] \hline \hline
(1,2) & $-$34   &  $-${\bf 12}
& {\bf 8} &  $-$2 & 0 & 0 \st\\[1ex] \hline
(1,4) & 0   & {\bf 0}
&  {\bf 0} & {0} & 0 & 0 \st\\[1ex] \hline
(2,4) & 0  &  {\bf  0} & {\bf  0} &  {\bf 0}
&  {0} &  0 \st\\[1ex] \hline
(1,6) & $-$156  &  $-${\bf  60}
&  {\bf  24} & $-${\bf12} & 4  & 0  \st\\[1ex] \hline
(2,6) & 0 &  {\bf 0} & {\bf 0}   & 
{\bf 0}  &  
{\bf 0} 
&  0 \st\\[1ex] \hline
(3,6) &  $-$672 &  $-${\bf 184} & {\bf 87}  & 
$-${\bf 24}
&  {\bf 10} & $-${\bf12}  \st\\[1ex] \hline
 \hline 
\end{tabular}
\caption{Values of $-B_6^{g_4}$ for the  $\ZZZ_4$-twined partition function of the $\ZZZ_2$ CHL model  
for different values of $Q^2$, $P^2$ and
$Q \cdot P$. The boldfaced entries are for charges
that satisfy the constraints (\ref{constraints_2}).} \label{table_Z2_Z4}
\end{center} }
\end{table}

\subsection{$\ZZZ_3~\times~\ZZZ_3$}
\label{sec:Z3Z3}

The $\ZZZ_3~\times~\ZZZ_3$ model is another example that can be dealt with in a straightforward manner. Values of $-B_6$ for the $\ZZZ_3$ CHL model (with no twining) 
for different values of $Q^2$, $P^2$ and
$Q \cdot P$ are given in table \ref{Z3_Z3_1}.  The Borcherds product formula \eqref{explift} with the index 3 weight 0 Jacobi form, 
\be 
\varphi = 2 \phi_{1}^{(4)} = 8 f_2^2 f_3^2 f_4^2
\ee
gives the $\ZZZ_3~\times~\ZZZ_3$ twisted-twining partition function \cite{Govindarajan:2019ezd}. The exponents in \eqref{explift} turn out to be $A = B = 1, C = 1/3$. This product can be programmed in \verb+Mathematica+. We find Fourier coefficients as listed in table \ref{Z3_Z3_2}. 
Using tables \ref{Z3_Z3_1} and \ref{Z3_Z3_2} it is easy to check that the expected positivity properties are satisfied. Once again, it is quite remarkable (almost a miracle, if you wish) that $-S_0, -S_1, -S_2$ all turn out to be positive integers.

\begin{table}[h!] {\small
\begin{center}\def\st{\vrule height 3ex width 0ex}
\begin{tabular}{|l|l|l|l|l|l|l|l|l|l|l|} \hline 
\backslashbox{$(Q^2,P^2)$}{$Q \cdot P$}
 &  $-$2 
& 0 & 1 & 2 & 3 & 4\st\\[1ex] \hline \hline
(2/3,2) & $-$1458   &  {\bf 540 }
&  {\bf 27} &  0 & 0 & 0 \st\\[1ex] \hline
(2/3,4) & $-$5616   & {\bf 3294}
&  {\bf 378 } & {0 } & 0 & 0 \st\\[1ex] \hline
(4/3,4) & $-$21496  &  {\bf 23008 } & {\bf  4912} &  {\bf 136 }
&  {0} & 0  \st\\[1ex] \hline
(2/3,6) & $-$18900  &  {\bf 16200 }
&  {\bf  2646} & {54} & 0  & 0  \st\\[1ex] \hline
(4/3,6) & $-$70524  &  {\bf 128706} & {\bf 37422}   & 
{\bf 2484}  &  
{ 6} 
&  {0} \st\\[1ex] \hline
(2,6) &  $-$208584 & 
{\bf 820404} &  {\bf 318267} & {\bf 37818 }  
&  {\bf 801 } & { 0 }  \st\\[1ex] \hline
 \hline 
\end{tabular}
\caption{Values of $-B_6$ for the $\ZZZ_3$ CHL model 
for different values of $Q^2$, $P^2$ and
$Q \cdot P$. The boldfaced entries are for charges
that satisfy the constraints to ensure that the index counts  microstates of a finite sized single centered black hole.  The constraints are given in \cite{1008.4209}.
We only
give the results for $3 Q^2 \le P^2$, since the index has a symmetry under $P^2\leftrightarrow
3 Q^2$. This table is identical to table 3 of \cite{1008.4209}. } \label{Z3_Z3_1}
\end{center} }
\end{table}
\begin{table}[h!] {\small
\begin{center}\def\st{\vrule height 3ex width 0ex}
\begin{tabular}{|l|l|l|l|l|l|l|l|l|l|l|} \hline 
\backslashbox{$(Q^2,P^2)$}{$Q \cdot P$}
 &  $-$2 
& 0 & 1 & 2 & 3 & 4\st\\[1ex] \hline \hline
(2/3,2) & 0   &  {\bf  0}
&  {\bf 0} &  0 & 0 & 0 \st\\[1ex] \hline
(2/3,4) & 0   & {\bf 0}
&  {\bf 0 } & {0 } & 0 & 0 \st\\[1ex] \hline
(4/3,4) & $-$124 &  $-${\bf  8} & $-${\bf 8 } &  {\bf  4}
&  {0} & 0  \st\\[1ex] \hline
(2/3,6) & 0  &  {\bf 0 }
&  {\bf 0 } & { 0} & 0  & 0  \st\\[1ex] \hline
(4/3,6) & 0  &  {\bf 0} & {\bf 0}   & 
{\bf 0}  &  
{ 0} 
&  {0} \st\\[1ex] \hline
(2,6) &  0 & 
{\bf 0} &  {\bf 0} & {\bf 0 }  
&  {\bf  0} & { 0 }  \st\\[1ex] \hline
 \hline 
\end{tabular}
\caption{Values of $-B_6^{g_3}$ for the $\ZZZ_3$-twined partition function of the $\ZZZ_3$ CHL model 
for different values of $Q^2$, $P^2$ and
$Q \cdot P$. The boldfaced entries are for charges
that satisfy the constraints to ensure that the index counts microstates of a finite size single centered black hole.  
We only
give the results for $3 Q^2 \le P^2$, the index has a symmetry under $P^2\leftrightarrow
3 Q^2$. } \label{Z3_Z3_2}
\end{center} }
\end{table}

We end this section with a comment about the asymptotic growth of Fourier coefficients and the corresponding logarithmic correction to the black hole entropy. In \cite{Belin:2016knb}, the authors identified a class of Siegel modular forms that could serve as ``candidates for other types of black holes'' by looking at the asymptotic growth of the Fourier coefficients of the various Siegel modular forms.   Two of the examples they identified are the  Borcherds lift \eqref{explift} of index 2 and 3, weight 0 Jacobi forms, as discussed above. The CHL interpretation of these modular forms is known (though, not well appreciated) as the
$\ZZZ_2~\times~\ZZZ_2$ and $\ZZZ_3~\times~\ZZZ_3$ twisted-twining partition functions, respectively \cite{1006.3472,Govindarajan:2019ezd}. Logarithmic corrections for the corresponding black holes on the gravity side have not been studied. We do expect the logarithmic corrections coefficients to match with the analysis of \cite{Belin:2016knb}. A proof of the equivalence of the various product representations of the twisted-twining partition functions for 
the $\ZZZ_2~\times~\ZZZ_2$ and $\ZZZ_3~\times~\ZZZ_3$ models has also not been written down.

\setcounter{equation}{0}

\section{Conclusions}
 
\label{sec:disc}

In this paper, we have studied indices counting the number of black hole microstates with definite eigenvalue under the $\ZZZ_N$ twining generator for a class of $\ZZZ_M$ CHL  models. Our study of course, forces the K3 moduli to lie in a subspace of the full moduli space, where such a  symmetry is geometrically realised. The number of black hole microstates in a $\ZZZ_M$ CHL model with definite eigenvalue under a $\ZZZ_N$ twining generator must be positive.  This leads to a specific prediction for the signs of certain linear combinations of Fourier coefficients 
 of Siegel modular forms. We explicitly tested these predictions for low charges. We studied these indices in a possible duality frames where the black holes do not admit more hair other than the fermionic zero modes associated to broken supersymmetries.  We also studied these indices for a sub-class of models in the D1-D5 duality frame.  In the D1-D5 duality frame, we computed the appropriate hair removed partition functions and showed the positivity of the appropriate Fourier coefficients for low charges.  We emphasise that ours is the first ever systematic study of the numerical computation of twined indices. Many of the subtle points that we have pointed out have not been  appreciated in the literature, e.g., the nature of the attractor chamber constraints on the charges.

For large charges, the twining indices for the twining generator of order $N$ are known to grow as~\cite{0911.1563},
\be 
\pm \exp\left[ \frac{S_{BH}}{N} \right], \label{S_BH_over_N}
\ee
where $S_{BH}$ is the entropy of a black hole carrying the same
set of charges. On the other hand the untwined indices grow as, 
\be 
+\exp \left[ S_{BH} \right].  \label{S_BH}
\ee
Numbers \eqref{S_BH_over_N} are exponentially small compared to \eqref{S_BH}. 
Clearly, the sum or difference of numbers \eqref{S_BH_over_N} and \eqref{S_BH}, as in \eqref{linear_comb}, will not change the positivity property of the large numbers \eqref{S_BH}. Thus, for large charges our results are nothing more than a consistency check. However, for low charges our results are fairly non-trivial. We have shown that in all the cases that we analysed, the positivity and integer property continues to hold.  These results are expected from the black hole side, but are quite intriguing from the modular form side, as linear combinations of Siegel Modular forms of different $\text{Sp}(2, \ZZZ)$ weights are involved.

Our results offer several opportunities for future research.  In a series of papers,  Chattopadhyaya and David~\cite{Chattopadhyaya:2017ews, Chattopadhyaya:2018xvg, Chattopadhyaya:2020yzl} have pointed out that the sign of the indices for  T$^4$ models violates the positivity conjecture of \cite{1008.4209}. 
In their most recent paper \cite{Chattopadhyaya:2020yzl}\footnote{There are some minor discrepancies in the $\ZZZ_2$ and $\ZZZ_5$ CHL tables in \cite{Chattopadhyaya:2020yzl}. We thank  A.~Chattopadhyaya and J.~David for confirming this. \label{footnote:AJ}}, they have proposed a ``tentative resolution'' of this puzzle. Their resolution is essentially based on the study of Fourier coefficients,  where they argue that 4 of the fermion zero modes for the $\frac{1}{4}$-BPS black holes for T$^4$ models are not hair modes.\footnote{We find this somewhat surprising. In the black hole hair removal program, the fermion zero modes are almost always taken to be hair modes. This is well-motivated, as shown in an earlier paper \cite{hep-th/9505116}.} A proper justification of this claim is still missing. Clearly, these models require further investigation. It will be interesting to explore the positivity property of the twisted-twining indices for the T$^4$ models. At the very least, such a study will provide a more refined version of the puzzle.

 We implemented Sen's product representation \eqref{twisted_twining_main} only for the 
 $\ZZZ_2~\times~\ZZZ_2$ 
 twisted-twining partition function. This representation is perhaps physically the most transparent, though fairly cumbersome to implement when it comes to extracting the Fourier coefficients.  We found it much easier to implement other representations, namely product of  genus two theta functions, and the Borcherds lift \eqref{explift}.   For other twisted-twining partition functions we only implemented the product of genus two theta functions representation and the Borcherds lift. It will be useful to confirm these results using Sen's product representation \eqref{twisted_twining_main}.\footnote{It will be useful to list $F^{(r,s; r',s')}(\tau, z)$ functions explicitly for a class of twisted-twining models in the future. These functions can  be used in  \eqref{twisted_twining_main} to provide additional   consistency checks on our results.} More broadly, it will be useful to work out proofs showing the equivalence of the various product representations of the twisted-twining partition functions. Some work in this direction has already been done in \cite{1312.0622}.  We note that, for twining partition functions (no twisting) these proofs have been recently completed~\cite{Govindarajan:2011em, Govindarajan:2020owu}.  We hope to return to some of the above problems in our future work.  
 
 \subsection*{Acknowledgements}
We thank Abhishek Chowdhury for collaboration at the very early stages of this project. We thank A.~Chattopadhyaya, J.~David,  A.~Sen, and R.~Volpato for email correspondence.
SS would like to thank IACS Kolkata for a fellowship  and hospitality where part of this work was done. The work of AV and PS was supported in part by the Max Planck Partnergroup ``Quantum Black Holes'' between CMI Chennai and AEI Potsdam and by a grant to CMI from the Infosys Foundation.

\bibliographystyle{jhep}

\bibliography{Draft_Apr_2022}

\end{document}